\documentclass[aps,prb,twocolumn,showpacs,amsmath,amssymb]{revtex4-2}

\usepackage{graphicx} 
\usepackage{dcolumn}  
\usepackage{bm}       
\usepackage{physics}
\usepackage{tikz}
\usetikzlibrary{arrows.meta}
\usepackage{float}
\usepackage{multirow}
\usepackage{amsmath}
\usepackage{mathtools}
\usepackage{url}

\usepackage{enumerate}
\usepackage{hyperref}

\hypersetup{
    colorlinks=true,
    linkcolor=blue,
    filecolor=magenta,
    urlcolor=blue,
    citecolor=blue,
}

\newcommand{\kket}[1]{ \left. \left| #1 \right> \hspace{-0.07cm} \right> } 
\newcommand{\bbra}[1]{ \left< \hspace{-0.07cm} \left< #1 \right| \right. }
\newcommand{\bbrakket}[2]{ \left. \left<  \hspace{-0.07cm} \left< #1 \right| #2 \right>\hspace{-0.07cm} \right> }

\begin{document}

\title{Free fermion models with an operator entanglement growing as \texorpdfstring{$O(\sqrt{t})$}{O(sqrt(t))}}

\author{Federico Tonetto and J\'er\^ome Dubail}

\affiliation{University of Strasbourg and CNRS, CESQ and ISIS (UMR 7006), 67000 Strasbourg, France}

\date{\today}

\begin{abstract}
We study the time evolution of the Operator Space Entanglement Entropy (OSEE) of one-dimensional number-conserving free fermions subject to a random space- and time-dependent evolution. It is known that, in free-fermion systems without randomness, the OSEE of \emph{local} operators grows at most \emph{logarithmically} in time. Here we show that, in the presence of randomness, the OSEE of local spin operators involving a Jordan-Wigner string grows as $\mathcal O(\sqrt t)$, for all values of the R\'enyi index. We establish this in two models: a Brownian Hamiltonian and a brick-wall matchgate circuit. We also consider the OSEE of the evolution operator, and we show that it also grows as $\mathcal O(\sqrt{t})$, in contrast with the linear growth generically expected without randomness. We analytically compute the average entropies of the two operators by mapping the quantum models to the symmetric simple exclusion process, relying on an assumption of self-averaging which we support numerically. We also consider two other types of randomness: random evolution in time that keeps spatial translation invariance, and quenched disorder. In the former case, we find that the OSEE also grows as $\sqrt{t}$. In the latter case, we find that the entropy grows as $\mathcal O( \log \log t)$ for Hamiltonian evolution both for local operators with a Jordan-Wigner string and for the evolution operator, whereas it remains bounded at all times in the brick-wall matchgate circuit.
\end{abstract}

\maketitle


\begin{figure*}[t]
    \centering
    \includegraphics[width=0.88\textwidth]{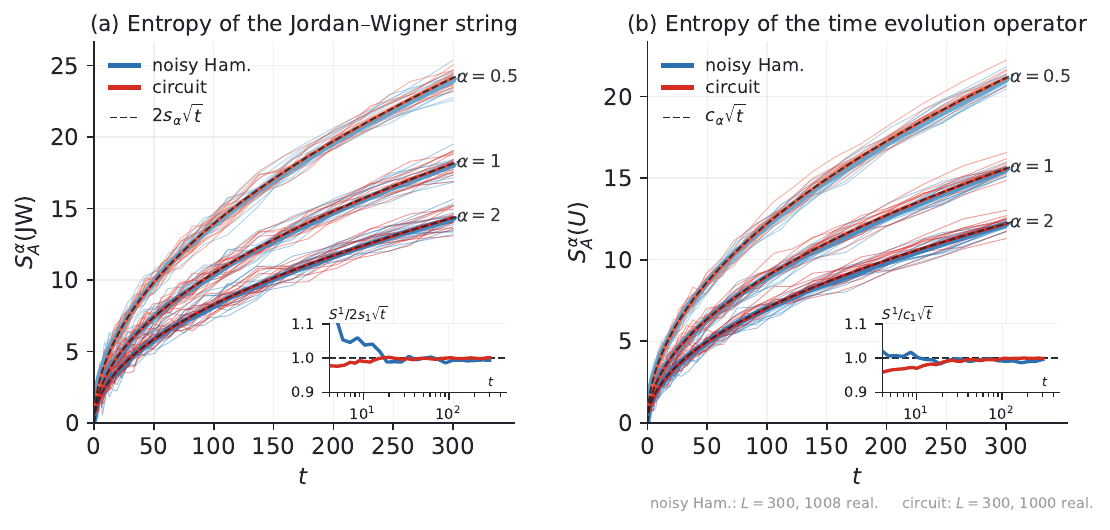}
    \caption{ \emph{Main result.} Time evolution of the OSEE of the two operators of interest, the Jordan--Wigner string with endpoint in the middle of the chain and the time evolution operator, under fully random spatio--temporal disorder, at $\alpha=\tfrac12,1,2$. The cut is in the middle of the chain: $A=\{-L/2+1,\cdots,0\}$, with $L$ being the length of the chain. We are plotting together the chain evolved with the noisy Hamiltonian (QSSEP) of Eq.~\eqref{eq:QSSEP_dH} and the random matchgate quantum circuit \eqref{eq:matchgate}.  The dashed curves correspond to the analytical values of the entropies. Faint lines represent single realizations; even if the two models are different microscopically, they agree on the values of the entropy. \emph{(a)} the OSEE of the Jordan--Wigner string, whose OSEE has the same time behavior of the OSEE of the local spin operators $X_j, Y_j$, against the closed--form amplitudes $s_\alpha$ of Eq.~\eqref{eq:Renyi}. The inset represents the fraction ${S^1}/{2s_1\sqrt{t}}$ in log scale, which approaches $1$ at large times. \emph{(b)} the OSEE of the time evolution operator, against the closed--form amplitudes $c_\alpha$ of Eq.~\eqref{eq:c_alpha}. The inset represents the fraction ${S^1}/{c_1\sqrt{t}}$ in log scale, which approaches $1$ at large times.}
    \label{fig:osee_full}
\end{figure*}

\section{Introduction}

    The operator space entanglement entropy (OSEE) \cite{Zanardi_2001,prosen2007operator,Prosen_OSEE_XY,Dubail_2017,Zhou_OSEE_evolution_operator_chaotic_systems} is often viewed as a quantifier of the `complexity' of an operator acting on the Hilbert space of a one-dimensional quantum system. Similarly to the entanglement entropy, which characterizes the approximability of quantum states by Matrix Product States \cite{Vestraete_2006,Hastings_2006, Vestraete_2008, Hartmann_2009, Cirac_2015}, the OSEE characterizes the approximability of operators by Matrix Product Operators (MPO), at least when the distance between them is measured with respect to the Hilbert-Schmidt norm. The idea is that, if the OSEE of an operator is large, then its approximation by an MPO requires a large bond dimension, so it is typically computationally hard to perform tasks with that operator on a classical computer~\cite{prosen2007operator,Prosen_OSEE_XY,znidaric2008complexity,Dubail_2017,Noh2020efficientclassical,rakovszky2022dissipation,wellnitz2022rise,dowling_2026_classicalsimulabilityoperatorentanglement}.
    

    The OSEE was introduced in 2001 by Zanardi in an attempt to quantify the amount of entanglement generated by a quantum operator $O$ acting on a bipartite Hilbert space $\mathcal{H}_A \otimes \mathcal{H}_B$~\cite{Zanardi_2001}. The definition mimics the one of the entanglement entropy for quantum states: one considers the Schmidt decomposition of the operator $O$, i.e. $O/\sqrt{\Tr (O^\dagger O)} =\sum_i \sqrt{\lambda_i} O_A^{(i)}\otimes O_{ B}^{(i)}$, with  $\Tr(O_A^{(i)\dagger}O_A^{(j)})=\delta_{ij}$. Then the R\'enyi-$\alpha$ OSEE is
    \begin{equation}
        S^\alpha_A(O) := \frac{1}{1-\alpha}\log\sum_i \lambda_i^\alpha.
        \label{eq:OSEE_def}
    \end{equation}
    Equivalently, the OSEE \eqref{eq:OSEE_def} is defined as the entanglement entropy of the vectorization of the operator $O$ (see e.g. Sec.~\ref{sec:jw} below). 
    
    In the context of quantum spin chains, the OSEE was first studied in 2007 by Prosen and Pi{\v{z}}orn  \cite{prosen2007operator,Prosen_OSEE_XY}, and since then it has been investigated in various one-dimensional systems, for several types of operators. The latter include density matrices~\cite{znidaric2008complexity,Dubail_2017,Noh2020efficientclassical,wellnitz2022rise,preisser2023comparing,rath2023entanglement,li2025highly,zhang2025operator}, time-evolution operators~\cite{Dubail_2017,Zhou_OSEE_evolution_operator_chaotic_systems,maccormack2021operator}, and local operators evolving in Heisenberg picture~\cite{prosen2007operator,Prosen_OSEE_XY,Muth_2011_numerics,Dubail_2017,Nahum_2018,Alba_2019,Bertini_2020,Bertini_2020_II,alba2021diffusion,kudler2021entanglement,medenjak2022operator,dowling2023scrambling,Murciano_2024,dowling2025bridging,alba2025more,Gopalakrishnan_2026,dowling_2026,dowling_2026_classicalsimulabilityoperatorentanglement,correr_2026,pozsgay2026growth}.
    
    The OSEE of local operators in Heisenberg picture, also called `local operator entanglement', appears to be a particularly remarkable quantity, as it is believed to be able to truly distinguish quantum many-body chaotic dynamics from integrable dynamics in quantum spin chains away from semiclassical limits. Specifically, 
    it is widely believed that the growth of the OSEE of local operators is at most logarithmic with time under integrable dynamics ~\cite{Alba_2019,Bertini_2020_II,pozsgay2026growth, alba2021diffusion,dowling2023scrambling} while it is linear in chaotic systems~\cite{Nahum_2018,Bertini_2020}, in agreement with an early conjecture about the scaling of the bond dimension needed to approximate these operators with MPOs formulated by Prosen and {\v{Z}}nidari{\v{c}}~\cite{Prosen_2007}. We note, however, that this dichotomy has been challenged recently by the discovery of a non-integrable model where the OSEE of local operators grows logarithmically \cite{TanProsen_2026}. (See also the note added at the end of this preprint.)

    \emph{OSEE in free fermion spin chains.}--- Refs.~\cite{prosen2007operator,Prosen_OSEE_XY,Dubail_2017} studied the OSEE of local operators in Heisenberg picture in spin chains that can be mapped to free fermions. These works focused on `clean' translation-invariant spin chains with time-independent Hamiltonians that map to quadratic Hamiltonians under a Jordan-Wigner transformation (for an introduction see e.g.~\cite{Tagliacozzo_2022_review}). The simplest example is the XX spin chain, with the Hamiltonian $H = \sum_{j} X_j X_{j+1} + Y_j Y_{j+1}$ where $X_j$, $Y_j$, $Z_j$ are the Pauli operators, which maps to $H \propto \sum_{j} c^\dagger_j c_{j+1} + c^\dagger_{j+1} c_{j}$ under
    \begin{eqnarray}
\label{eq:JW_transform}
\nonumber  X_j &=& \mathrm{JW}_{j-1} ( c^\dagger_j + c_j) \\
\nonumber  Y_j &=& \mathrm{JW}_{j-1}  \, i ( c_j^\dagger- c_j) \\
\nonumber  Z_j &=& 1-2c^\dagger_j c_j \, .
\end{eqnarray}
 Here the operator noted `${\rm JW}_j$' is the Jordan-Wigner (JW) string going from the left end of the chain ($-\infty$ if the chain is infinite) up to site $j$,
    \begin{equation}
    \label{eq:JW_string}
        \mathrm{JW}_j := \prod_{k \leq j} \big(1-2c^\dagger_k c_k \big).
    \end{equation}
    Refs.~\cite{prosen2007operator,Prosen_OSEE_XY,Dubail_2017} found that the OSEE of local spin operators (e.g. $X_j$, $Y_j$, $Z_j$, $X_j X_{j+1}$, $X_j Y_{j+1}$, $Z_j  Y_{j+1} X_{j+2}$, etc.) crucially depends on whether or not they involve a JW string when they are represented in terms of the fermions.
    
    Local spin operators whose fermionic representation does not include a JW string, such as $Z_j$ in the XX chain, all have a bounded OSEE. In contrast, operators with a JW string, such as $X_j$ or $Y_j$ in the XX chain, have an OSEE that grows as
\begin{equation}
    S^\alpha_A(O) = \frac{c}{6}\Big(1+\frac{1}{\alpha}\Big)\log t + \mathcal{O} (1),
    \label{eq:temporal_log}
\end{equation}
    at late times. (Here $c$ is an integer number that depends on the dispersion relation of the free fermions~\cite{Prosen_OSEE_XY}; for the XX spin chain  $c=1$.) These results for free fermions are in line with the aforementioned widespread belief that the OSEE of local operators grows at most logarithmically in systems with integrable dynamics.
    
    \emph{Question motivating this work.}--- We revisit the question of the growth of the OSEE in spin chains that map to free fermions and therefore are `integrable'. In contrast with Refs.~\cite{prosen2007operator,Prosen_OSEE_XY,Dubail_2017}, we consider random time-evolution operators, with randomness either in space or in time, or both. 
    Like in the clean case, the local spin operators whose fermion representation does not include a JW string (e.g. $Z_j$ above) all have a bounded OSEE, a fact already pointed out in Ref.~\cite{Essler_2022_QSSEP_useful_formulas} in the context of the `Quantum Symmetric Simple Exclusion Process'~\cite{TonyJin_2019_openQSSEP}.

    Here we ask what happens to the operators with the JW string (e.g. $X_j$ or $Y_j$ above), i.e. the ones whose OSEE grows as (\ref{eq:temporal_log}) in the clean case. How does their OSEE behave in random free fermion models? Does it grow logarithmically, hence reflecting the underlying `integrability' or `freeness' of the dynamics? Or is it linear, reflecting the fact that the randomness of the evolution makes the system more `chaotic'?

      \emph{Main result.}--- The answer turns out to be  somewhere in between. In this work we find that, in random free fermion models with spatio-temporal disorder, local spin operators $O$ with a JW string have an OSEE that grows, on average, as
      \begin{equation}
        \label{eq:main_result}
        S^\alpha_A (O) = 2 s_\alpha \sqrt{t} + o(\sqrt t)
      \end{equation}
        instead of the logarithmic growth (\ref{eq:temporal_log}). (The coefficient $s_\alpha$ will be studied in detail in Section~\ref{sec:analytic}.)
        
        A typical situation is shown in Fig.~\ref{fig:osee_full}(a), where we compute the OSEE of the JW string operator (\ref{eq:JW_string}) for different random models (a Brownian Hamiltonian evolution and a brick-wall circuit, presented in detail in Section~\ref{sec:models}). Fig.~\ref{fig:osee_full} also shows that the $\sqrt t$ growth holds for each single realization, not only on average. Following Ref.~\cite{Dubail_2017}, we focus on the OSEE of the JW string operator ${\rm JW}_j$ itself, as opposed to particular cases of operators involving the string, like $X_j$ and $Y_j$ in the example above . Indeed, since all these operators consist in the string ${\rm JW}_j$ multiplied by a finite number of local fermionic modes (of the form $c_j^\dagger$, $c_j$, $c^\dagger_j c^\dagger_{j+1} c^\dagger_{j+2}$, $c^\dagger_j c_{j+1} c_{j+2}$ etc.), and the latter always keep a finite Schmidt rank under free fermion evolution, the large-time asymptotic behavior of their OSEE is always identical to the one of the string ${\rm JW}_j$. Thus, it is sufficient to study the quantity
        \begin{equation}
            \label{eq:OSEE_string}
            S^\alpha_A ( U^\dagger(t) {\rm JW}_j  U(t) )
        \end{equation}
        where $U(t)$ is the evolution operator of the model. This is the main quantity of interest throughout the paper, and our main result is that it grows according to Eq.~(\ref{eq:main_result}).

     \emph{Other result of this work: OSEE of evolution operator.}--- The OSEE of the evolution operator $U(t)$ was investigated in Refs.~\cite{Dubail_2017,Zhou_OSEE_evolution_operator_chaotic_systems} (see also Ref.~\cite{maccormack2021operator} about closely related quantities). In particular, 
    it was found that the OSEE of the time evolution operator generically grows linearly with time regardless of the integrability of the system \cite{Dubail_2017}, with the only known exception being many-body localized systems \cite{Zhou_OSEE_evolution_operator_chaotic_systems}. Here we revisit this question in random free fermion models. We compute the OSEE of the evolution operator, and we find that
    \begin{equation}
        \label{eq:main_result2}
        S^\alpha_A( U(t) ) \, = \, c_\alpha \sqrt{t} + o(\sqrt{t}) \, ,
    \end{equation}
    in stark contrast with the results of Refs.~\cite{Dubail_2017,Zhou_OSEE_evolution_operator_chaotic_systems}. A typical situation is shown in Fig.~\ref{fig:osee_full}(b).

     Technically, the calculation of $S^\alpha_A( U(t) )$ in free fermion systems is intimately related to the one of $ S^\alpha_A ( U^\dagger(t) {\rm JW}_j  U(t) )$ (as explained in Section~\ref{sec:jw}), and this is our main motivation for presenting our results for both quantities simultaneously.

\emph{Organization of the paper.}--- We start by reviewing the procedure for vectorizing free fermionic systems in  Section~\ref{sec:jw}. In particular we show that the two objects of interest, the Jordan--Wigner string and the evolution operator, map to free-fermionic vectorized states, and we express their OSEE as two different functions of the spectrum of a common correlation matrix. This is the main technical point that is used to obtain all other results in the paper.
In Section~\ref{sec:models} we define the free-fermionic models that we investigate. Those include continuous-time Hamiltonian and discrete-time circuit models. In Sec.~\ref{sec:numerics} we present our numerical results for the two classes of models and for the different types of disorder we have investigated. Finally, in Section~\ref{sec:analytic} we present our analytical derivation of Eqs.~(\ref{eq:main_result})-(\ref{eq:main_result2}) for the case of fully random disorder, i.e. disorder both in space and in time. We conclude in Section~\ref{sec:conclusion}.
    
\section{Vectorization in free fermionic systems}
\label{sec:jw}

In this section we discuss in full generality the vectorization in free fermionic systems with number conservation. We apply the vectorization to the Jordan--Wigner string and to the evolution operator. We exploit the fact that both the vectorized Jordan--Wigner string and the vectorized evolution operator are Gaussian states, and we derive the key formula~\eqref{eq:twofunctions} which expresses their OSEE in terms of the eigenvalues of a correlation matrix of dimension $O(L \times L)$ where $L$ is the system length. Formula~\eqref{eq:twofunctions} is used extensively in the rest of the paper.

\subsection{Vectorization and fermions}

\emph{Definition.}---Given a Hilbert space $\mathcal{H}$, vectorization maps operators $O\in\text{End}(\mathcal{H})$ to states $\kket{O}\in\mathcal{H}\otimes \tilde{\mathcal{H}}$, where $\tilde{\mathcal{H}}$ is a copy of the Hilbert space. We use a double-bracket notation in order to avoid confusion with vectors $\left| v \right>$ in the original Hilbert space $\mathcal{H}$. The map is not unique \nocite{2026_Essler_vectorization}\footnote{Vectorizations different from Eq.~\eqref{def: vectorization} are possible. For example Eq.~(26) of \cite{2026_Essler_vectorization}. However that vectorization has not the property \eqref{eq:isometry}, hence Eq.~\eqref{eq:OSEE=EE} does not hold. Of course, this is not a problem for \cite{2026_Essler_vectorization} because they were not interested in computing the OSEE and their vectorization is a smart choice for treating their situation.}; for our purposes (computing the OSEE) it is convenient to require that the map $ \text{End}(\mathcal{H}) \owns O \mapsto \kket{O} \in \mathcal{H} \otimes \mathcal{\tilde H}$ is an isometry,
\begin{equation}
\label{eq:isometry}
    \Tr(O_1^\dagger O_2 ) = \bbrakket{O_1}{O_2}.
\end{equation}
Concretely, the map we choose is
\begin{equation}
    \label{def: vectorization}
     O:=\sum_{i,j} O_{ij} \ket{i}\bra{j} \;\longmapsto\; \kket{O} = \sum_{i,j} O_{ij} \ket{i} \tilde{\ket{j}}.
\end{equation}
In particular, the identity is sent to the (unnormalized) maximally entangled state
\begin{equation}
    \label{eq: vectorization of identity}
    \kket{\mathbb{I}} = \sum_{j} \ket{j} \ket{\tilde{j}},
\end{equation}
which we will use throughout as a reference state. Every operator is obtained from it by multiplication,
\begin{equation}
    \label{eq: vectorization of O}
    \kket{O} = (O \otimes \tilde{\mathbb{I}}) \kket{\mathbb{I}} .
\end{equation}
Note that, equivalently,
\begin{equation}
   \kket{O} = (\mathbb{I}\otimes \tilde{O}^T )  \kket{\mathbb{I}} .
   \label{eq:relation_OOT}
\end{equation}

The vectorization~\eqref{def: vectorization} ensures that the operator Schmidt decomposition entering in Eq.~\eqref{eq:OSEE_def} is identical to the Schmidt decomposition of $\kket{O}$, hence the OSEE of $O$ is equal to the usual Rényi entanglement entropy of the state $\kket{O}$,
\begin{equation}
    S^\alpha_A(O)=S^\alpha_A(\kket{O}) .
    \label{eq:OSEE=EE}
\end{equation}

\emph{Vectorization and fermions.}---
We consider a one-dimensional chain of length $L$ of spinless fermions $c_j,c^\dagger_j$, $\{c_j,c^\dagger_k\}=\delta_{jk}$. We take $L$ even, and we label the fermions from $j = -\frac{L}{2}+1$ to $\frac{L}{2}$. The Hilbert space $\mathcal{H}$ is the fermionic Fock space generated by the action of the creation operators on the vacuum.

To define the vectorization, we need a copy $\tilde{\mathcal{H}}$ of that Hilbert space. For this, we introduce new fermion creation operators $\tilde{c}^\dagger_j$ such that $\tilde{\mathcal{H}}$ is the Fock space generated by these modes. We take the modes $c^\dagger_j$ and $\tilde{c}^\dagger_j$ to be mutually anti-commuting. Then the operator space ${\rm End} (\mathcal{H}) \simeq \mathcal{H} \otimes \tilde{\mathcal{H}}$ is the Fock space generated by all the $c_j^\dagger$'s and $\tilde{c}^\dagger_j$'s. The reference state \eqref{eq: vectorization of identity} is
\begin{equation}
    \kket{\mathbb{I}} = \prod_{j=-L/2+1}^{L/2} \left(1+c^\dagger_j \tilde{c}^\dagger_j  \right)\kket{0},
    \label{eq:Omega_ferm}
\end{equation}
where $\tilde{c}^\dagger_j$ creates a fermion in the auxiliary copy and $\kket{0}$ is annihilated by both the $c_j$'s and $\tilde c_j$'s. 

Acting with $c^\dagger_j$ and $c_j$ on the reference state \eqref{eq:Omega_ferm} gives the two relations
\begin{equation}
    c_j\kket{\mathbb{I}}=\tilde c^\dagger_j\kket{\mathbb{I}},
    \qquad
    c^\dagger_j\kket{\mathbb{I}}=-\tilde c_j\kket{\mathbb{I}} .
    \label{eq:Omega_rel}
\end{equation}
Note that the latter equality is in contradiction with (\ref{eq:relation_OOT}). This is because we are now working with fermions, so the tensor product in Eq.~(\ref{eq: vectorization of O})-(\ref{eq:relation_OOT}) should be understood as a $\mathbb{Z}_2$-graded tensor product. In other words, the correct way to write these formulas for fermionic operators is
\begin{equation}
    \label{eq:vectorization_O_Z2}
    \kket{O} = (O \otimes \tilde{\mathbb{I}}) \kket{\mathbb{I}} , \qquad \kket{O} = (-1)^F (\mathbb{I}\otimes \tilde{O}^T )  \kket{\mathbb{I}} ,
\end{equation}
where $(-1)^F$ is the fermion parity operator.

\subsection{Vectorization of Gaussian operators and of Jordan-Wigner strings}

\emph{Vectorized Gaussian operator.}--- In the following Sections we will consider time-evolution operators that are Gaussian and preserve the total number of fermions. Such operators are of the form
\begin{equation}
\label{eq:Hamiltonian_general}
    U := \exp \left( i \ \sum_{i,j} J_{ij} c^\dagger_i c_j \right) \, ,
\end{equation}
with $J$ an $L \times L$ Hermitian matrix. Crucially, the vectorization of (\ref{eq:Hamiltonian_general}) is a Gaussian state. Indeed, from the definitions we see that
\begin{align*}
    \kket{U} &=  (U \otimes \mathbb{I}) \kket{\mathbb{I}} \\
    &=  (U \otimes \mathbb{I}) \exp (\sum_j c^\dagger_j \tilde{c}^\dagger_j ) \kket{0} \\
    &= \exp (\sum_j U c^\dagger_j U^\dagger   \tilde{c}^\dagger_j ) ( U\otimes \mathbb{I}) \kket{0} .
\end{align*}
Because of number conservation, we have $( U \otimes \mathbb{I}) \kket{0}=\kket{0}$. Moreover, the Gaussian unitary operator $U$ acts by conjugation on the fermion modes as
\begin{equation}
    U^\dagger c_j U = \sum_{k} u_{jk} c_k, \qquad u := \exp(i J) \, .
    \label{eq:gate_general}
\end{equation}
Therefore,
\begin{equation}
\label{eq:vec of U final}
\begin{aligned}
    \kket{U}=\exp\Big(\sum_{j,k} u_{jk} c^\dagger_j\tilde{c}^\dagger_k\Big)\kket{0} \, .
\end{aligned}
\end{equation}
This is a BCS-like state (Bardeen-Cooper-Schrieffer), with a total number of fermions $c^\dagger$ and $\tilde{c}^\dagger$ that is not fixed. Thus, computing the OSEE of $U$ boils down to evaluating the entanglement entropy of a BCS state. We will exploit this fact below.

\emph{The vectorized Jordan--Wigner string.}--- The other operator on which we will be focusing in this paper is the Jordan--Wigner string (\ref{eq:JW_string}). 

\begin{figure}[t]
    \centering
    \begin{tikzpicture}[
        occ/.style={circle,draw=#1,fill=#1,inner sep=2.1pt},
        emp/.style={circle,draw=#1,fill=white,line width=0.8pt,inner sep=2.1pt},
        pair/.style={gray!55,line width=0.7pt},
      ]
      \colorlet{pcol}{blue!62!black}
      \colorlet{acol}{orange!88!black}
      \def\dx{0.95}

      \begin{scope}
        \draw[pcol!40,line width=1.0pt] (-0.3,0.72) -- ({7*\dx+0.3},0.72);
        \draw[acol!40,line width=1.0pt] (-0.3,0)    -- ({7*\dx+0.3},0);
        \foreach \i in {0,...,3}{
          \node[occ=pcol] at (\i*\dx,0.72) {};
          \node[occ=acol] at (\i*\dx,0) {};}
        \foreach \i in {4,...,7}{
          \node[emp=pcol] at (\i*\dx,0.72) {};
          \node[emp=acol] at (\i*\dx,0) {};}
        \draw[gray!70,dashed,line width=0.7pt]
          ({3.5*\dx},-0.52) -- ({3.5*\dx},1.22);
        \node[gray!55!black,font=\tiny,anchor=south] at ({3.5*\dx},1.18) {cut};
        \node[pcol,font=\scriptsize,anchor=east] at (-0.42,0.72) {$b_j$};
        \node[acol,font=\scriptsize,anchor=east] at (-0.42,0) {$d_j$};
        \node[pcol,font=\tiny,anchor=west] at ({7*\dx+0.34},0.72) {$u^\dagger$};
        \node[acol,font=\tiny,anchor=west] at ({7*\dx+0.34},0) {$u^{T}$};
        \node[font=\footnotesize,anchor=west] at (-0.42,1.55)
          {$\kket{\mathrm{JW}}$: one domain wall per species of particle};
      \end{scope}
    \end{tikzpicture}
    \caption{ Vectorization of the Jordan--Wigner string. In the combinations $b_j,d_j$ of Eq.~\eqref{eq:bd} the reference state is the vacuum and the time evolution \eqref{eq:L_bd} is decoupled, propagating with $u^\dagger$ and $u^{T}$. The Jordan--Wigner string is the Fock state with all the modes to the left of the cut occupied on both species of particles, Eq.~\eqref{eq:JW0}, so that its OSEE twice the entanglement entropy of a single domain wall \eqref{eq:factor2}.}
    \label{fig:ladder}
\end{figure}
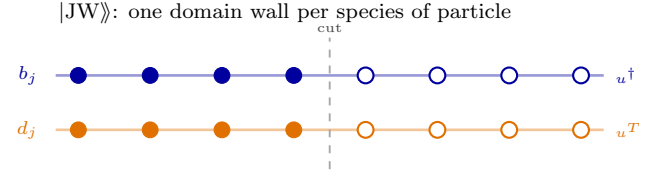

The vectorization of the Jordan--Wigner string follows from Eqs.~\eqref{eq: vectorization of O} and \eqref{eq:Omega_rel},
\begin{equation}
    \label{eq: vec of JW}
    \kket{\text{JW}_0 }=\prod_{j\le0}\big(1-2c^\dagger_j \tilde{c}^\dagger_j \big)\kket{\mathbb{I}} \, . 
\end{equation}
It is convenient to do a Bogoliubov transformation \cite{Dubail_2017,Murciano_2024} and to introduce the modes
\begin{equation}
    b_j := \frac{c_j-\tilde{c}^\dagger_j}{\sqrt2},\qquad 
    d_j := \frac{c^\dagger_j+\tilde{c}_j}{\sqrt2},
    \label{eq:bd}
\end{equation}
which obey the canonical anticommutation relations  $\{b_j,b^\dagger_k\}=\{d_j,d^\dagger_k\}=\delta_{jk}$ and $\{b_j,d_k\}=\{b_j,d^\dagger_k\}=0$ and
annihilate the reference state,
\begin{equation}
    b_j\kket{\mathbb{I}}=d_j\kket{\mathbb{I}}=0 .
    \label{eq:bd_vacuum}
\end{equation}
Under that Bogoliubov transformation, the Jordan--Wigner string \eqref{eq: vec of JW} becomes
\begin{equation}
    \label{eq:JW0}
    \begin{aligned}
    \kket{\rm{JW}_0} = \prod_{j\le0}\big (-b^\dagger_j d^\dagger_j\big)\kket{\mathbb{I}} \, .
    \end{aligned}
\end{equation}
This is a tensor product of two domain-wall states~\cite{Dubail_2017,Murciano_2024}
\begin{equation}
\begin{aligned}
    &\kket{\rm{JW}_0} \propto \ket{\rm{DW}_0 }_b \otimes \ket{\rm{DW}_0 }_d, \\
    &\ket{\rm{DW}_0 }_b:= \prod_{j \leq 0 } b^\dagger_j \kket{\mathbb{I}}, \qquad \ket{\rm{DW}_0}_d:= \prod_{j \leq 0 } d^\dagger_j \kket{\mathbb{I}},
\end{aligned}
\end{equation}
up to an unimportant sign. This is illustrated in Fig.~\ref{fig:ladder}. 

In the rest of the paper we will be interested in the time-evolved Jordan-Wigner string. Therefore we need to be able to conjugate the operator ${\rm JW}_0$ by a Gaussian unitary operator $U$ of the form \eqref{eq:Hamiltonian_general}. It turns out this is easily done with the Bogoliubov transformation (\ref{eq:bd}). To see this, observe that, for any operator $O$, it follows from the definition that $\kket{ U^\dagger O U } = U^\dagger O U \otimes \tilde{\mathbb{I}} \kket{ \mathbb{I} } =  U^\dagger O \otimes  \tilde{U}^T  \kket{ \mathbb{I} } =  \mathcal{U} (O \otimes \tilde{\mathbb{I}} ) \kket{ \mathbb{I} } = \mathcal{U} \kket{ O }$ where
\begin{equation}
    \mathcal{U} := U^\dagger \otimes \tilde{U}^T , \qquad \tilde{U} := \exp (i \sum_{ij} J_{ij} \tilde{c}^\dagger_i  \tilde{c}_j ) .
\end{equation}
The superoperator $\mathcal{U}$ acts independently on the $b$ and $d$ modes (\ref{eq:bd}),
\begin{equation}
    \mathcal{U}^\dagger b_j \mathcal{U} = \sum_{k} [u^\dagger]_{jk} b_k, \qquad \mathcal{U}^\dagger d_j \mathcal{U} = \sum_{k} [u^T]_{jk} d_k \, .
    \label{eq:L_bd}
\end{equation}
Therefore the vectorization of the $U$-conjugated Jordan-Wigner string, $\kket{ U^\dagger {\rm JW}_0 U} = \mathcal{U} \kket{{\rm JW}_0}$, remains a product state of two domain wall states, one for the $b$ modes and the other for the $d$ modes,
\begin{equation}
    \begin{aligned}
    \mathcal{U} \kket{\rm{JW}_0} &= e^{-i\sum_{ij} J_{ij} b^\dagger_i b_j } \ket{\rm{DW}_0}_b \otimes e^{i\sum_{ij} J_{ij}^* d^\dagger_i d_j }\ket{\rm{DW}_0}_d \\
    &=  (U^\dagger \left| {\rm DW}_0 \right> )_b \otimes (U^T \left| {\rm DW}_0 \right> )_d \, .
    \end{aligned}
\end{equation}
Consequently, the OSEE of the JW string conjugated by a Gaussian unitary operator $U$ is twice the entanglement entropy of a single domain-wall state acted upon by $U$~\cite{Dubail_2017,Murciano_2024},
\begin{equation}
    S^\alpha_A\big(U^\dagger{\rm JW}_0 U\big)=2\,S^\alpha_A\big({U}^\dagger\ket{{\rm DW}_0}_b\big)  \, .
    \label{eq:factor2}
\end{equation}

\subsection{OSEE of Gaussian operators and of Jordan--Wigner strings}

From now on we always consider the bipartition $A \cup B$ of the chain
\begin{equation}
    A=\{ -\frac{L}{2}+1, \dots, -1, 0 \} \quad \text{and} \quad  B=\{1, 2, \dots, \frac{L}{2}\}    .
    \label{eq:bipatition}
\end{equation}
We are interested in the OSEE of the Gaussian unitary operator $U$ and in the OSEE of the Jordan-Wigner string ${\rm JW}_0$ conjugated by $U$. To calculate those, we apply Peschel's formula to the vectorization discussed in the previous subsection. Our main result is formulas \eqref{eq:falpha}-\eqref{eq:twofunctions}, which express the OSEE of both operators in terms of the spectrum of the correlation matrix \eqref{eq:Gdef} restricted to the subsystem $B$.

\emph{Reduced correlation matrix and its spectrum.---}
Consider the $L \times L$ correlation matrix
\begin{equation}
    G^{U}_{jk} := \bra{\mathrm{DW}} U b^\dagger_k b_j U^\dagger \ket{\mathrm{DW}} .
    \label{eq:Gdef}
\end{equation}
Note that $G^U$ is Hermitian, and that it is equal to
\begin{equation}
    \label{eq:Gevolution}
    G^U = u^\dagger\, \text{diag}( \{ \theta(-j) \}_{-\frac{L}{2}+1 \leq j \leq \frac{L}{2}} )  \, u 
\end{equation}
where $\theta$ is the Heaviside step function with $\theta(0) =1$, and $u$ is the $L \times L$ unitary matrix $ u = e^{i J}$, see Eq.~(\ref{eq:gate_general}).

The reduced correlation matrix $G^U_{BB}$ is the block with row and column indices corresponding to subsystem $B$ of the chain,
\begin{equation}
    G^U \, = \, \left( \begin{array}{c|c}
         G^U_{AA} &  G^U_{AB}  \\ \hline 
         G^U_{BA} & G^U_{BB}
    \end{array} \right) \, .
\end{equation}
More generally, we will use the notation $M_{AA}$, $M_{AB}$, etc. for blocks of a matrix $M$ with row and column indices in the corresponding subsystem. Restricting Eq.~\eqref{eq:Gevolution} to the subsytem $B$ yields
\begin{equation}
   G^U_{BB} = (u_{AB})^{\dagger} u_{AB} \, .
    \label{eq:GB}
\end{equation}
In what follows, we will use the spectrum of $G^U_{BB}$, and the spectrum of $u_{AA} (u_{AA})^\dagger$,
\begin{align}
    \label{def:spec_of_G_B}
    \{ \tau_k \}_{1 \leq k \leq L/2} = {\rm spec} ( \, G^U_{BB} ) \, , \\
    \{\lambda_k\}_{1 \leq k \leq L/2} = {\rm spec} ( \, u_{AA} (u_{AA})^\dagger ) \, .
    \label{def:singular_values_u_AA}
\end{align}
The latter is not independent from the spectrum $\{ \tau_k \}$. Let us show why. The unitarity of $U$ implies
\begin{equation}
    u_{AA}^{\phantom{\dagger}}(u_{AA})^{\dagger}+u_{AB}^{\phantom{\dagger}} (u_{AB})^{\dagger}=\mathbb{I}_A.
    \label{eq:unitarity}
\end{equation}
Thus the two matrices $u_{AA}^{\phantom{\dagger}}(u_{AA})^{\dagger}$ and $u_{AB}^{\phantom{\dagger}} (u_{AB})^{\dagger}$ are simultaneously diagonalizable.
The $L_B \times L_B$ matrix $(u_{AB})^{\dagger} u_{AB}$ and the $L_A \times L_A$ matrix $u_{AB}^{\phantom{\dagger}} (u_{AB})^{\dagger}$ (here with $L_A = L_B = L/2$) have the same non-vanishing eigenvalues $\{\tau_k\}$; here since $L_A = L_B$ their whole spectrum is exactly the same.
Thus Eq.~\eqref{eq:unitarity} implies
\begin{equation}
   \lambda_k + \tau_k =1, \qquad \text{for} \quad k=1,...,\tfrac{L}{2} \, .
    \label{eq:lambdatau}
\end{equation}

\emph{Peschel formula for the OSEE of the JW string.}--- Let us recall the Peschel formula for the Rényi entanglement entropy of free fermion states with fixed particle number~ \cite{Peschel_2003, Peschel_2009}. The formula gives the entanglement entropy of the bipartition $A \cup B$ in terms of the spectrum $\{\tau_k\}$ of the reduced correlation matrix,
\begin{equation}
    S^\alpha_A=\sum_k h_\alpha(\tau_k),
    \qquad
    h_\alpha(\tau):=\frac{1}{1-\alpha}\log\big[\tau^\alpha+(1-\tau)^\alpha\big].
    \label{eq:peschel}
\end{equation}
The von Neumann entropy is obtained for $\alpha \rightarrow 1$, with $h_1(\tau)=-\tau\log\tau-(1-\tau)\log(1-\tau)$. Note that $h_\alpha(\tau)=h_\alpha(1-\tau)$ and $h_\alpha(0)=h_\alpha(1)=0$.

Applying the Peschel formula \eqref{eq:peschel} to the correlation matrix \eqref{eq:Gdef}, together with the relation between the entanglement entropy of the domain wall and the OSEE of the Jordan--Wigner string \eqref{eq:factor2}, we get
\begin{equation}
    S^\alpha_A(U^\dagger {\rm JW}_0 U) = 2 \sum_{k} h_\alpha (\tau_k) \, .
    \label{eq:OSEE_JW}
\end{equation}

\emph{Peschel formula for Gaussian unitary operator.}--- Since the state $\kket{U}$ is a BCS-like state, with no fixed number of particles, see Eq.~\eqref{eq:vec of U final},  we need to apply a formula for the entanglement entropy that differs from the one of the previous paragraph, that is also due to Peschel~\cite{Peschel_2003,Peschel_2009}. The formula reads
\begin{equation}
    S^\alpha_A  =  \sum_k h_\alpha (s_k)
\end{equation}
where $h_\alpha$ is the same function as above and
\begin{equation}
    \{ s_k^\pm \}_{-L/2+1 \leq k \leq 0} = {\rm spec} (M_{AA})
\end{equation}
where $M_{AA}$ is the restriction to subsystem $A$ of the $2L \times 2L$ correlation matrix $M$ with blocks ($n,m = -L/2+1,\dots , L/2$)
\begin{equation}
\label{def:correlation_matrix_M}
    M_{nm}:= 
    \begin{pmatrix}
        \langle c_n c_m^\dagger \rangle & \langle c_n \tilde c_m \rangle \\
        \langle \tilde c^\dagger_n c_m^\dagger \rangle & \langle \tilde c^\dagger_n \tilde c_m \rangle
    \end{pmatrix}.
\end{equation}
The expectation value $\langle \cdot \rangle$ is taken with respect to the normalized state
\begin{equation}
    \langle \cdot \rangle := \frac{\bbra{U} \cdot \kket{U}}{\bbrakket{U}{U}} =\frac{\bbra{U} \cdot \kket{U}}{2^L} \, .
\end{equation}

We now relate the spectrum $\{s_k^\pm\}$ of the reduced correlation matrix $M_{AA}$ to the spectrum $\{\lambda_k\}$ of the matrix $u_{AA} (u_{AA})^\dagger$. 

In fact, the correlation matrix $M$ Eq.~\eqref{def:correlation_matrix_M} can be rewritten as
\begin{equation}
    M_{AA} = \frac{1}{2} \begin{pmatrix} \mathbb{I} &- u_{AA}\\ -u_{AA}^\dagger& \mathbb{I} \end{pmatrix} .
    \label{eq:MX}
\end{equation}
This follows from the unitarity of $U$ together with 
\begin{equation}
\begin{aligned}
    &\begin{aligned}
        2^{-L}\bbra{\mathbb{I}}c^\dagger_j c_k \kket{\mathbb{I}} &= \frac{2^{-L}}{2} \bbra{\mathbb{I}} \left(b^\dagger_j +d_j \right)\left(b_k +d^\dagger_k \right) \kket{\mathbb{I}} \\
        &= \frac{2^{-L}}{2} \bbra{\mathbb{I}}d_j d^\dagger_k \kket{\mathbb{I}} = \frac{\delta_{jk}}{2},
    \end{aligned} \\
    &2^{-L}\bbra{\mathbb{I}}\tilde c^\dagger_j \tilde c_k \kket{\mathbb{I}} = 2^{-L}\bbra{\mathbb{I}}\tilde c_j c_k \kket{\mathbb{I}} = \frac{\delta_{jk}}{2}, \\
    &2^{-L}\bbra{\mathbb{I}} c_j \tilde c_k \kket{\mathbb{I}} = -\frac{\delta_{jk}}{2}.
\end{aligned}
\label{eq:<I|c^+c|>}
\end{equation}
The eigenvalues of $M_{AA}$ are then given by
\begin{equation}
    s_k^\pm = \frac{1\pm \sqrt{\lambda_k} }{2}.
    \label{eq:Uspectrum}
\end{equation}
Finally, we obtain the desired relation between the spectrum of $M$ \eqref{eq:Uspectrum} and the eigenvalues of the correlation matrix of the domain wall \eqref{def:spec_of_G_B}
\begin{equation}
    s_k^\pm = \frac{1\pm \sqrt{1-\tau_k }}{2}.
    \label{eq:spectrum_relation_U_and_DW}
\end{equation}
Since $h_\alpha(s)=h_\alpha(1-s)$, the $\pm$ pairs in Eq.~\eqref{eq:spectrum_relation_U_and_DW} contribute equally, and Eq.~\eqref{eq:peschel} gives 
\begin{equation}
    S^\alpha_A\big(U\big)= 2 \sum_k f_\alpha (\tau_k),
\end{equation}
where 
\begin{equation}
    f_\alpha(\tau):=h_\alpha\!\Big(\frac{1+\sqrt{1-\tau}}{2}\Big) .
    \label{eq:falpha}
\end{equation}

\emph{Summary.}--- We have obtained two formulas for the OSEE of the Jordan--Wigner string and of a Gaussian unitary operator as functions of the spectrum $\{\tau_k\}$ \eqref{def:spec_of_G_B} of the reduced correlation matrix $G^U_{BB}$,
\begin{equation}
    \begin{aligned}
    S^\alpha_A(U^\dagger {\rm JW} U) &= 2 \sum_{k} \frac{1}{1-\alpha}\log\big(\tau_k^\alpha+(1-\tau_k)^\alpha\big), \\
    S^\alpha_A\big(U\big) &= 2 \sum_k \frac{1}{1-\alpha} \log\Big\{ \left(\frac{1+\sqrt{1-\tau_k}}{2} \right)^\alpha \\
    & \qquad \qquad \qquad \quad  +\left(\frac{1-\sqrt{1-\tau_k}}{2}\right)^\alpha\Big\}.
    \end{aligned}
    \label{eq:twofunctions}
\end{equation}
We stress that the factors $2$ in both formulas appeared for different reasons. For the Jordan--Wigner string, the factor $2$ in Eq.~(\ref{eq:twofunctions}) is a consequence of the OSEE being twice the entanglement entropy of a domain-wall state, see Eq.~\eqref{eq:factor2}, while for the Gaussian unitary operator $U$ it is due to the fact that, when we plug the entanglement spectrum \eqref{eq:Uspectrum} inside the Peschel's formula \eqref{eq:peschel}, the two values of $s_k^\pm = (1\pm \sqrt{\lambda_k})/2$ contribute equally.

Therefore the whole computation of the OSEE boils down to evaluating the eigenvalues $\{\tau_k\}$ of the reduced correlation matrix $G^U_{BB}$ in the domain-wall state \eqref{eq:Gdef}. This is what we do in the rest of the paper.

\section{The models}
\label{sec:models}
In this Section, we introduce the free fermion models in which we want to study the OSEE. As we just explained in Sec.~\ref{sec:jw}, this is done through analyzing the melting of a domain-wall state in these models. First, we specify the two types of models that we consider: a Hamiltonian model with continuous time evolution and a quantum circuit with discrete time evolution. Later we describe the types of disorder we will consider.

\subsection{Brownian Hamiltonian: the QSSEP}
The free fermionic Hamiltonian model with number conservation that we consider is a noisy nearest-neighbor hopping Hamiltonian with couplings that are both space- and time-dependent. This model is known under the name of Quantum Symmetric Simple Exclusion Process (QSSEP). It was introduced in Ref.~\cite{TonyJin_2019} and has attracted significant attention in the past years \cite{TonyJin_2019_openQSSEP, Piroli_2021_entanglementQSSEP, Essler_2022_QSSEP_useful_formulas, Hruza_2023_QSSEP_free_probability, Hruza_2023_entanglementlatetimelimitQSSEP, barraquand_2025_introductionquantumexclusionprocesses, JoaoCosta_2026, StefanoScopa_2026_universalclassicalquantumfluctuations}. The model is defined by the Hamiltonian increment
\begin{equation}
\begin{aligned}
    dH^t := \sum_{j}\left( dW^{t}_j\, c^\dagger_{j+1} c_j + d\overline{W}^{t}_j\, c^\dagger_{j} c_{j+1}\right), \\ 
    \mathbb{E}[dW^{t}_j]=0, \mathbb{E}[dW^{t}_j dW^t_k]=0  \quad  \text{and} \quad 
    \mathbb{E}[dW^{t}_j\, d\overline{W}^{t}_k] = \delta_{jk}\, dt, 
\end{aligned}
    \label{eq:QSSEP_dH}
\end{equation}
where the $dW^{t}_j$ are complex It\^o increments (white noise) and $\mathbb{E}$ is the noise average.

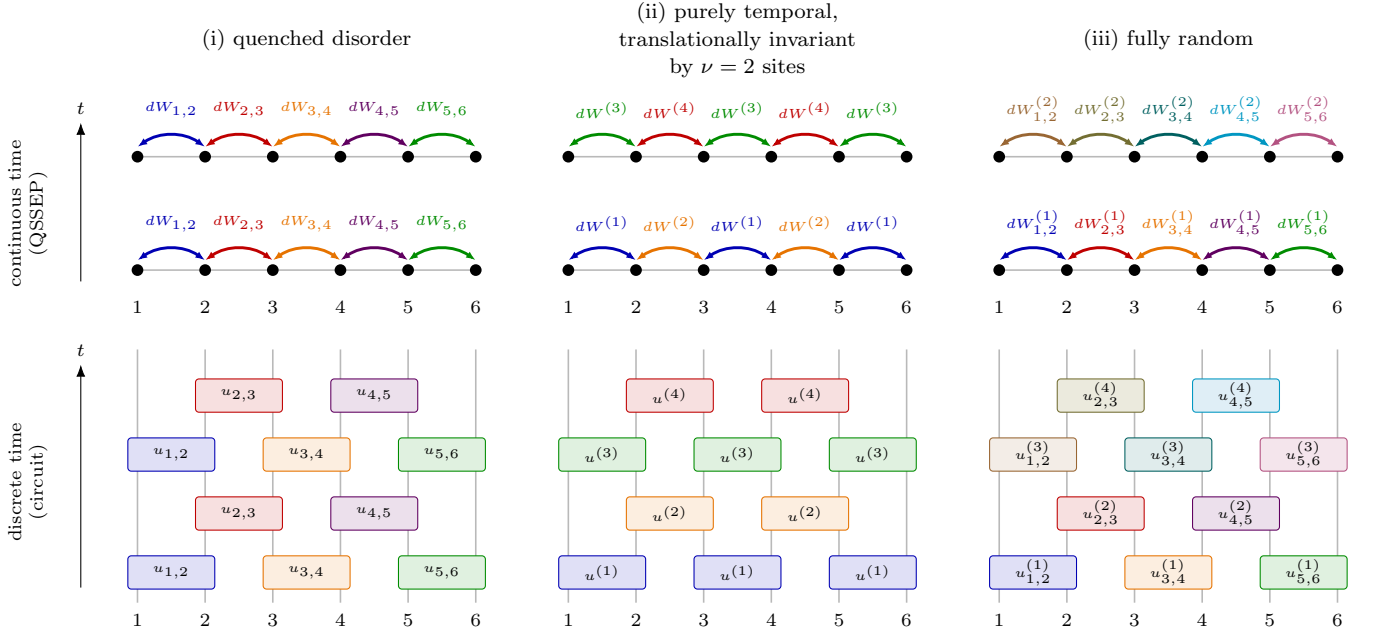
\begin{figure*}[!htb]
    \centering
    
    \newcommand{\qssepBase}{%
      \draw[wire] (0,0) -- (5*\dxw,0);
      \draw[wire] (0,\dyq) -- (5*\dxw,\dyq);
      \foreach \i in {0,...,5}{\fill (\i*\dxw,0) circle (2.2pt);
                               \fill (\i*\dxw,\dyq) circle (2.2pt);}
      \foreach \i/\lab in {0/1,1/2,2/3,3/4,4/5,5/6}
          {\node[below,font=\scriptsize] at (\i*\dxw,-0.30) {$\lab$};}
    }
    
    \newcommand{\qssepArc}[4]{
      \draw[#1,line width=1.1pt,{Latex[length=1.3mm]}-{Latex[length=1.3mm]}]
            (#2*\dxw,#4+0.13) to[bend left=35] (#3*\dxw,#4+0.13);
    }
   
    \newcommand{\circuitWires}{%
      \foreach \i in {0,...,5}{\draw[wire] (\i*\dxw,-0.40) -- (\i*\dxw,2.95);}
      \foreach \i/\lab in {0/1,1/2,2/3,3/4,4/5,5/6}
          {\node[below,font=\scriptsize] at (\i*\dxw,-0.43) {$\lab$};}
    }
    \begin{tikzpicture}[
        wire/.style={gray!55,line width=0.6pt},
        gt/.style={rounded corners=1.3pt,minimum width=1.15cm,
                   minimum height=0.44cm,inner xsep=0pt,inner ysep=0pt,
                   font=\tiny},
        hdr/.style={font=\footnotesize},
        tag/.style={font=\scriptsize},
        rowlab/.style={font=\scriptsize,rotate=90,align=center}
    ]
      
      \colorlet{cA}{blue!70!black}   \colorlet{cB}{orange!90!black}
      \colorlet{cC}{green!55!black}  \colorlet{cD}{red!75!black}
      \colorlet{cE}{violet!75!black} \colorlet{cF}{brown!80!black}
      \colorlet{cG}{teal!75!black}   \colorlet{cH}{magenta!70!black}
      \colorlet{cI}{olive!65!black}  \colorlet{cJ}{cyan!75!black}

      \def\dxw{0.895}
      \def\dyl{0.78}
      \def\dyq{1.5}
      \def\colB{5.70cm}
      \def\colC{11.40cm}

      \begin{scope}[xshift=0cm,yshift=4cm]
        \qssepBase
        \qssepArc{cA}{0}{1}{0}      \qssepArc{cD}{1}{2}{0}      \qssepArc{cB}{2}{3}{0}
        \qssepArc{cE}{3}{4}{0}      \qssepArc{cC}{4}{5}{0}
        \qssepArc{cA}{0}{1}{\dyq}   \qssepArc{cD}{1}{2}{\dyq}   \qssepArc{cB}{2}{3}{\dyq}
        \qssepArc{cE}{3}{4}{\dyq}   \qssepArc{cC}{4}{5}{\dyq}
        \node[cA,font=\tiny] at (0.5*\dxw,0.62)       {$dW_{1,2}$};
        \node[cD,font=\tiny] at (1.5*\dxw,0.62)       {$dW_{2,3}$};
        \node[cB,font=\tiny] at (2.5*\dxw,0.62)       {$dW_{3,4}$};
        \node[cE,font=\tiny] at (3.5*\dxw,0.62)       {$dW_{4,5}$};
        \node[cC,font=\tiny] at (4.5*\dxw,0.62)       {$dW_{5,6}$};
        \node[cA,font=\tiny] at (0.5*\dxw,\dyq+0.62)  {$dW_{1,2}$};
        \node[cD,font=\tiny] at (1.5*\dxw,\dyq+0.62)  {$dW_{2,3}$};
        \node[cB,font=\tiny] at (2.5*\dxw,\dyq+0.62)  {$dW_{3,4}$};
        \node[cE,font=\tiny] at (3.5*\dxw,\dyq+0.62)  {$dW_{4,5}$};
        \node[cC,font=\tiny] at (4.5*\dxw,\dyq+0.62)  {$dW_{5,6}$};

        \node[hdr]  at (2.5*\dxw,3.05) {(i) quenched disorder};
        \draw[-{Latex[length=1.6mm]}] (-0.75,-0.15) -- (-0.75,1.95)
              node[above,font=\scriptsize] {$t$};
        \node[rowlab] at (-1.45,0.75) {\shortstack{continuous time\\ (QSSEP)}};
      \end{scope}
      
      \begin{scope}[xshift=\colB,yshift=4cm]
        \qssepBase
        \qssepArc{cA}{0}{1}{0}      \qssepArc{cB}{1}{2}{0}      \qssepArc{cA}{2}{3}{0}
        \qssepArc{cB}{3}{4}{0}      \qssepArc{cA}{4}{5}{0}
        \qssepArc{cC}{0}{1}{\dyq}   \qssepArc{cD}{1}{2}{\dyq}   \qssepArc{cC}{2}{3}{\dyq}
        \qssepArc{cD}{3}{4}{\dyq}   \qssepArc{cC}{4}{5}{\dyq}
        \node[cA,font=\tiny] at (0.5*\dxw,0.62)       {$dW^{(1)}$};
        \node[cB,font=\tiny] at (1.5*\dxw,0.62)       {$dW^{(2)}$};
        \node[cA,font=\tiny] at (2.5*\dxw,0.62)       {$dW^{(1)}$};
        \node[cB,font=\tiny] at (3.5*\dxw,0.62)       {$dW^{(2)}$};
        \node[cA,font=\tiny] at (4.5*\dxw,0.62)       {$dW^{(1)}$};
        \node[cC,font=\tiny] at (0.5*\dxw,\dyq+0.62)  {$dW^{(3)}$};
        \node[cD,font=\tiny] at (1.5*\dxw,\dyq+0.62)  {$dW^{(4)}$};
        \node[cC,font=\tiny] at (2.5*\dxw,\dyq+0.62)  {$dW^{(3)}$};
        \node[cD,font=\tiny] at (3.5*\dxw,\dyq+0.62)  {$dW^{(4)}$};
        \node[cC,font=\tiny] at (4.5*\dxw,\dyq+0.62)  {$dW^{(3)}$};

        \node[hdr,align=center] at (2.5*\dxw,3.05)
            {\shortstack{(ii) purely temporal,\\ translationally invariant\\ by $\nu=2$ sites}};
      \end{scope}
      
      \begin{scope}[xshift=\colC,yshift=4cm]
        \qssepBase
        \qssepArc{cA}{0}{1}{0}      \qssepArc{cD}{1}{2}{0}      \qssepArc{cB}{2}{3}{0}
        \qssepArc{cE}{3}{4}{0}      \qssepArc{cC}{4}{5}{0}
        \qssepArc{cF}{0}{1}{\dyq}   \qssepArc{cI}{1}{2}{\dyq}   \qssepArc{cG}{2}{3}{\dyq}
        \qssepArc{cJ}{3}{4}{\dyq}   \qssepArc{cH}{4}{5}{\dyq}
        \node[cA,font=\tiny] at (0.5*\dxw,0.62)       {$dW^{(1)}_{1,2}$};
        \node[cD,font=\tiny] at (1.5*\dxw,0.62)       {$dW^{(1)}_{2,3}$};
        \node[cB,font=\tiny] at (2.5*\dxw,0.62)       {$dW^{(1)}_{3,4}$};
        \node[cE,font=\tiny] at (3.5*\dxw,0.62)       {$dW^{(1)}_{4,5}$};
        \node[cC,font=\tiny] at (4.5*\dxw,0.62)       {$dW^{(1)}_{5,6}$};
        \node[cF,font=\tiny] at (0.5*\dxw,\dyq+0.62)  {$dW^{(2)}_{1,2}$};
        \node[cI,font=\tiny] at (1.5*\dxw,\dyq+0.62)  {$dW^{(2)}_{2,3}$};
        \node[cG,font=\tiny] at (2.5*\dxw,\dyq+0.62)  {$dW^{(2)}_{3,4}$};
        \node[cJ,font=\tiny] at (3.5*\dxw,\dyq+0.62)  {$dW^{(2)}_{4,5}$};
        \node[cH,font=\tiny] at (4.5*\dxw,\dyq+0.62)  {$dW^{(2)}_{5,6}$};

        \node[hdr] at (2.5*\dxw,3.05) {(iii) fully random};
      \end{scope}

      \begin{scope}[xshift=0cm,yshift=0cm]
        \circuitWires
        \node[gt,draw=cA,fill=cA!12] at (0.5*\dxw,0)      {$u_{1,2}$};
        \node[gt,draw=cB,fill=cB!12] at (2.5*\dxw,0)      {$u_{3,4}$};
        \node[gt,draw=cC,fill=cC!12] at (4.5*\dxw,0)      {$u_{5,6}$};
        \node[gt,draw=cD,fill=cD!12] at (1.5*\dxw,\dyl)   {$u_{2,3}$};
        \node[gt,draw=cE,fill=cE!12] at (3.5*\dxw,\dyl)   {$u_{4,5}$};
        \node[gt,draw=cA,fill=cA!12] at (0.5*\dxw,2*\dyl) {$u_{1,2}$};
        \node[gt,draw=cB,fill=cB!12] at (2.5*\dxw,2*\dyl) {$u_{3,4}$};
        \node[gt,draw=cC,fill=cC!12] at (4.5*\dxw,2*\dyl) {$u_{5,6}$};
        \node[gt,draw=cD,fill=cD!12] at (1.5*\dxw,3*\dyl) {$u_{2,3}$};
        \node[gt,draw=cE,fill=cE!12] at (3.5*\dxw,3*\dyl) {$u_{4,5}$};
        \draw[-{Latex[length=1.6mm]}] (-0.75,-0.20) -- (-0.75,2.75)
              node[above,font=\scriptsize] {$t$};
        \node[rowlab] at (-1.45,1.20) {\shortstack{discrete time\\ (circuit)}};
      \end{scope}
      
      \begin{scope}[xshift=\colB,yshift=0cm]
        \circuitWires
        \node[gt,draw=cA,fill=cA!12] at (0.5*\dxw,0)      {$u^{(1)}$};
        \node[gt,draw=cA,fill=cA!12] at (2.5*\dxw,0)      {$u^{(1)}$};
        \node[gt,draw=cA,fill=cA!12] at (4.5*\dxw,0)      {$u^{(1)}$};
        \node[gt,draw=cB,fill=cB!12] at (1.5*\dxw,\dyl)   {$u^{(2)}$};
        \node[gt,draw=cB,fill=cB!12] at (3.5*\dxw,\dyl)   {$u^{(2)}$};
        \node[gt,draw=cC,fill=cC!12] at (0.5*\dxw,2*\dyl) {$u^{(3)}$};
        \node[gt,draw=cC,fill=cC!12] at (2.5*\dxw,2*\dyl) {$u^{(3)}$};
        \node[gt,draw=cC,fill=cC!12] at (4.5*\dxw,2*\dyl) {$u^{(3)}$};
        \node[gt,draw=cD,fill=cD!12] at (1.5*\dxw,3*\dyl) {$u^{(4)}$};
        \node[gt,draw=cD,fill=cD!12] at (3.5*\dxw,3*\dyl) {$u^{(4)}$};
      \end{scope}
      
      \begin{scope}[xshift=\colC,yshift=0cm]
        \circuitWires
        \node[gt,draw=cA,fill=cA!12] at (0.5*\dxw,0)      {$u^{(1)}_{1,2}$};
        \node[gt,draw=cB,fill=cB!12] at (2.5*\dxw,0)      {$u^{(1)}_{3,4}$};
        \node[gt,draw=cC,fill=cC!12] at (4.5*\dxw,0)      {$u^{(1)}_{5,6}$};
        \node[gt,draw=cD,fill=cD!12] at (1.5*\dxw,\dyl)   {$u^{(2)}_{2,3}$};
        \node[gt,draw=cE,fill=cE!12] at (3.5*\dxw,\dyl)   {$u^{(2)}_{4,5}$};
        \node[gt,draw=cF,fill=cF!12] at (0.5*\dxw,2*\dyl) {$u^{(3)}_{1,2}$};
        \node[gt,draw=cG,fill=cG!12] at (2.5*\dxw,2*\dyl) {$u^{(3)}_{3,4}$};
        \node[gt,draw=cH,fill=cH!12] at (4.5*\dxw,2*\dyl) {$u^{(3)}_{5,6}$};
        \node[gt,draw=cI,fill=cI!12] at (1.5*\dxw,3*\dyl) {$u^{(4)}_{2,3}$};
        \node[gt,draw=cJ,fill=cJ!12] at (3.5*\dxw,3*\dyl) {$u^{(4)}_{4,5}$};
      \end{scope}
    \end{tikzpicture}
    \caption{The two models and the three types of disorder, on the same six sites and for two consecutive time steps. \emph{Top row:} the noisy Hamiltonian of Eq.~\eqref{eq:QSSEP_dH}, each arc is the hopping $dW^{t}_j$ of that bond. \emph{Bottom row:} the brick--wall matchgate circuit, one Haar random $U(2)$ gate per bond and per layer, odd and even bonds alternating. In both rows {equal colours denote the same random variable}. (i) \emph{Quenched disorder}: the randomness depends on the bond but not on time, $dW^{t}_j\to W_{j}dt$ for the Hamiltonian and the same layer repeated at every step for the circuit. (ii) \emph{Purely temporal disorder}: the randomness is redrawn at every instant but at each instant there are only two independent It\^o increments, so the chain is translation invariant by $\nu=2$ sites; in the circuit one gate per layer is shared by all its bonds. (iii) \emph{Fully random disorder}: an independent increment $dW^t_j$, resp.\ an independent gate, on every bond and at every time.}
    \label{fig:models}
\end{figure*}

\subsection{Discrete time: Haar-random matchgate circuits}
The second setting is a brick--wall random \emph{Gaussian} circuit. The evolution is discrete. The evolution operator between the step $t$ and $t+1$ is $U^{(t)}=U^{(t)}_{\rm even}U^{(t)}_{\rm odd}$ with $U^{(t)}_{\rm odd}=\bigotimes_{j}  U^{(t)}_{2j-1,2j}$ and $U^{(t)}_{\rm even}=\bigotimes_{j} U^{(t)}_{2j,2j+1}$. Random brick-wall circuits have been extensively studied in the literature \cite{Nahum_2017_randomcircuitsentanglementgrowth,Nahum_2018_randomcircuitsoperatorspreading,Zhou_2020_randomcircuitsentanglementmembrane,Pollmann_2018_randomcircuits,Pollmann_2018_randomcircuitsconservationlaws,Rakovszky_2019_randomcircuitsconservationlaws}. Here, crucially, we restrict our unitary gates $U^{(t)}_{j,j+1}$ to Gaussian gates, also known as \emph{matchgates}. Brick--wall circuits of random matchgates have been considered previously in Refs.~\cite{Ribeirp2021, lastres2026geometryfreefermioncommutants}.

Gaussian two-qubit gates, or matchgates, are of the form (in the computational basis)
\begin{equation}
    U(v,u)_{j,j+1}:=\begin{pmatrix}
    v_{11} & 0 & 0 & v_{12}\\
    0 & u_{11} & u_{12} & 0\\
    0 & u_{21} & u_{22} & 0\\
    v_{21} & 0 & 0 & v_{22}
    \end{pmatrix}_{j,j+1}
    \label{eq:matchgate}
\end{equation}
where $u$ and $v$ are $2\times 2$ matrices with
\begin{equation}
        \det v=\det u ,
        \label{eq:determinant_constrain}
\end{equation}
i.e.\ gates acting as $v$ on the even--parity subspace $\{\ket{00},\ket{11}\}$ and as $u$ on the odd--parity one $\{\ket{01},\ket{10}\}$, with equal determinants. It was shown that circuits of such gates acting on nearest neighbours can be simulated classically in polynomial time \cite{Valiant_2002}. The reason was later identified by \cite{TerhalDiVincenzo_2002,Knill_2001}: the Jordan--Wigner transformations \eqref{eq:JW_string} applied to a nearest--neighbour matchgate give the exponential of a \emph{quadratic} fermionic Hamiltonian, $U_{j,j+1}=e^{-iH_{j,j+1}}$ with $H_{j,j+1}$ a fermion bilinear. So, a matchgate circuit is a system of non--interacting fermions (see also \cite{JozsaMiyake_2008} for a systematic account).

Number conservation is the statement $v_{12}=v_{21}=0$. After fixing the irrelevant overall phase so that $v_{11}=1$, the determinat constrain~\eqref{eq:determinant_constrain} gives $v_{22}=\det u$ in Eq.~\eqref{eq:matchgate}. Thus, the gate then does nothing to $\ket{00}$, multiplies $\ket{11}$ by $\det u$, and acts as the $U(2)$ matrix $u$ on the two single-particle states. Under the application of the layers $U^{(t)}_{\rm even}$ and $U^{(t)}_{\rm odd}$ the fermionic operators transform linearly among themselves,
\begin{equation}
    {U^{(t)}}^\dagger_{\text{odd}} c_i U^{(t)}_{\text{odd}} = \sum_{j} [u^{(t)}_{\text{odd}}]_{ij} c_j  \, ,
\end{equation}
with the same equation for ${{U}^{(t)}}^\dagger_{\rm even}$. The single-particle evolution matrices $u^{(t)}_{\text{odd}}$ and $u^{(t)}_{\text{even}}$ have the following block-diagonal structures,
\begin{equation}
    u^{(t)}_{\text{odd}} = 
    \left(
    \begin{array}{cc|cc|cc|cc}
        \multicolumn{2}{c|}{\multirow{2}{*}{$u^{(t)}_{1,2}$}} & 0 & 0 & 
        \multicolumn{2}{c|}{\cdots} & 0 & 0 \\
        \multicolumn{2}{c|}{} & 0 & 0 & \multicolumn{2}{c|}{\cdots} & 0 & 0 \\
        \hline
        0 & 0 & \multicolumn{2}{c|}{\multirow{2}{*}{$u^{(t)}_{3,4}$}} & \multicolumn{2}{c|}{\cdots} & 0 & 0 \\
        0 & 0 & \multicolumn{2}{c|}{} & \multicolumn{2}{c|}{\cdots} & 0 & 0 \\
        \hline
        \multirow{2}{*}{$\vdots$} & \multirow{2}{*}{$\vdots$} & \multirow{2}{*}{$\vdots$} & \multirow{2}{*}{$\vdots$} & \ddots & & \multirow{2}{*}{$\vdots$} & \multirow{2}{*}{$\vdots$} \\
        & & & & & \ddots & & \\
        \hline
        0 & 0 & 0 & 0 & \multicolumn{2}{c|}{\cdots} & \multicolumn{2}{c}{\multirow{2}{*}{$u_{L-1,L}$}} \\
        0 & 0 & 0 & 0 & \multicolumn{2}{c|}{\cdots} & \multicolumn{2}{c}{}
    \end{array}
    \right)
\end{equation}
and
\begin{equation}
    u^{(t)}_{\text{even}} = 
    \left(
    \begin{array}{c|cc|cc|cc|c}
        1 & 0 & 0 & 0 & 0 & \multicolumn{2}{c|}{\cdots} & 0 \\
        \hline
        0 & \multicolumn{2}{c|}{\multirow{2}{*}{$u^{(t)}_{2,3}$}} & 0 & 0 & \multicolumn{2}{c|}{\cdots} & 0 \\
        0 & \multicolumn{2}{c|}{} & 0 & 0 & \multicolumn{2}{c|}{\cdots} & 0 \\
        \hline
        0 & 0 & 0 & \multicolumn{2}{c|}{\multirow{2}{*}{$u^{(t)}_{4,5}$}} & \multicolumn{2}{c|}{\cdots} & 0 \\
        0 & 0 & 0 & \multicolumn{2}{c|}{} & \multicolumn{2}{c|}{\cdots} & 0 \\
        \hline
        \multirow{2}{*}{$\vdots$} & \multirow{2}{*}{$\vdots$} & \multirow{2}{*}{$\vdots$} & \multirow{2}{*}{$\vdots$} & \multirow{2}{*}{$\vdots$} & \ddots & & \multirow{2}{*}{$\vdots$} \\
        & & & & & & \ddots & \\
        \hline
        0 & 0 & 0 & 0 & 0 & \multicolumn{2}{c|}{\cdots} & 1
    \end{array}
    \right).
\end{equation}

We sample the $2\times 2$ matrices $ u^{(t)}_{j,j+1}$ from the Haar measure on $U(2)$ \cite{Collins_2006,Mele_2024}. We stress that this is different from the the Haar-random $U(4)$ gates of generic chaotic random circuits~\cite{Nahum_2017_randomcircuitsentanglementgrowth,Nahum_2018_randomcircuitsoperatorspreading,Zhou_2020_randomcircuitsentanglementmembrane,Pollmann_2018_randomcircuits,Pollmann_2018_randomcircuitsconservationlaws,Rakovszky_2019_randomcircuitsconservationlaws}: rather than randomizing over the full two--site unitary group, one restricts to the largest subgroup that leaves the free, Gaussian structure intact, together with the symmetry of the system, and randomizes within it.

\subsection{Types of disorder considered in this work}
\label{sec:types of randomness}

In this work we will consider three types of randomness as in Ref.~\cite{Ribeirp2021}, represented in Fig.~\ref{fig:models}.

\begin{enumerate}[(i)]
\item \emph{Quenched (static) disorder}: randomness in space but not in time. For the random Hamiltonian of Eq.~\eqref{eq:QSSEP_dH}, this correspond to the case in which the It\^o increments are frozen in time $dW^t_j= W_j dt$ where $W_j$ is time independent. In the random circuit we fix the gates once and we repeat the same layer at every step.

\item \emph{Random evolution with translation invariance}: we let the system evolve with a random time evolution operator that is invariant under translations by $\nu$ sites (see  Fig.~\ref{fig:models}(ii) for an illustration of the $\nu = 2$ case). Concretely, for the circuit evolution, a single gate is drawn at each time step and shared by all the
bonds of that layer, so that the circuit is invariant under two–site translations. For the Hamiltonian evolution, a translation-invariant version of the noisy Hamiltonian~\eqref{eq:QSSEP_dH} has been proposed in Refs.~\cite{Alba_nuQSSEP_intro, Alba_nuQSSEP_mpemba,Alba_nuQSSEP_GHD}. There the authors considered Eq.~\eqref{eq:QSSEP_dH} with $\nu$ independent It\^o increments such that 
\begin{equation}
    \mathbb E [dW^t_j \ d\overline{W}_k^t] = \delta_{j \equiv k \, \text{mod } \nu} dt.
\end{equation}
This is the model we consider here. In particular, the most relevant case for our purposes is the $\nu=2$ case, which is closest to the translation-invariant circuit case, see Fig.~\ref{fig:models}(ii). We also discuss the $\nu=1$ case in the App.~\ref{app:nu=1}, which we stress has no counterpart in the circuit, because a brick--wall circuit cannot be translationally invariant by just one site.

\item \emph{Fully random}: we have disorder both in space and time. This means that both the It\^o increments $dW^t_j$ and the $2\times2$ unitary matrices $u^{(t)}_{j,j+1}$ are drawn independently at each position and time.

\end{enumerate}

\section{Numerical results}
\label{sec:numerics}

In this Section we present our numerical results for the OSEE of the time-evolved Jordan-Wigner \eqref{eq:JW_string} and for the time-evolution operator $U(t)$ itself in the models presented in Section~\ref{sec:models}. Here the random evolution operator is:
\begin{itemize}
    \item $U(t) = \mathcal{T} \exp \left( - i \int^t_0 d H^t \right)$ for the Brownian Hamiltonian evolution,
    \item $U(t) = U^{(t)}U^{(t-1)}\cdots U^{(1)}$ for the random circuit evolution.
\end{itemize}
In both cases this is a Gaussian unitary operator, so all the results derived in Section~\ref{sec:jw} apply. In particular, for each realization of disorder, the numerical simulation consists propagating the correlation matrix $G^U$, see Eq.~\eqref{eq:Gevolution}, and in diagonalizing its restriction to the subsystem $B =\{1,\dots,\frac{L}{2}\}$. From the spectrum \eqref{def:spec_of_G_B} of the correlation matrix, we can determine the OSEE of both operators using Eqs.~\eqref{eq:twofunctions}.


The matchgates~\eqref{eq:matchgate} of our circuits are sampled according to the $U(2)$ Haar measure using the unique QR decomposition \cite{mezzadri2007}. For the Brownian Hamiltonian evolution, the It\^o increments $dW^t_j$ used in the simulation are
\begin{equation}
    dW^t_j = \mathcal{N}_1(0,\sigma^2) + i\mathcal{N}_2(0,\sigma^2),
\end{equation}
where $\mathcal{N}_{1,2}(0,\sigma^2)$ are two Gaussians independently identically distributed with $0$ average and variance $\sigma^2= dt/2$ with $dt$ being a small time step taken to be $dt=0.05$.

In the following we present our results for the three types of disorder (i), (ii), (iii) introduced in Subsection~\ref{sec:types of randomness}, for both continuous-time and discrete-time evolution.

\begin{figure*}[!htb]
    \centering
    \includegraphics[width=0.88\textwidth]{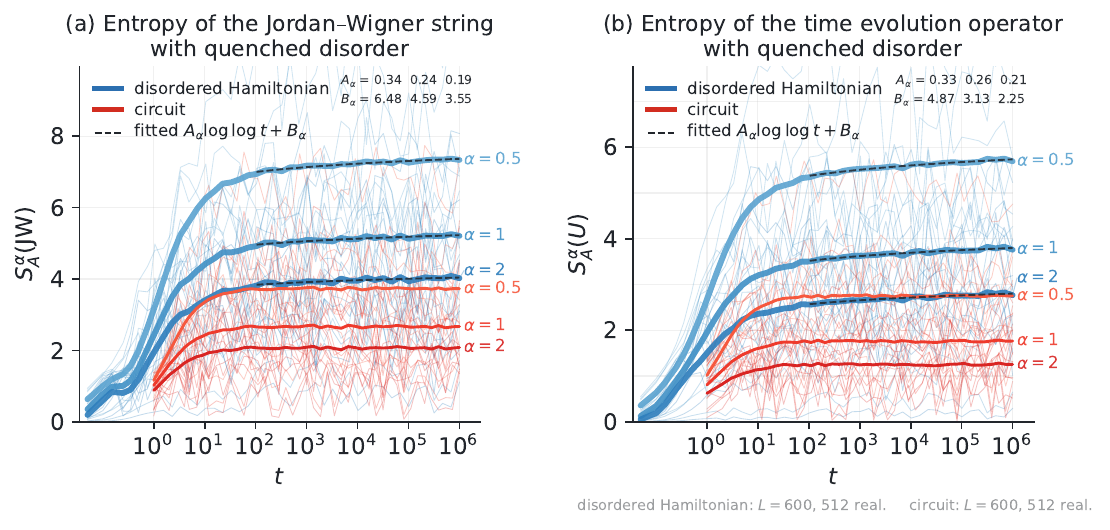}
    \caption{OSEE of the Jordan--Wigner string (left) and of the time
    evolution operator (right) with quenched disorder,
    Sec.~\ref{sec:types of randomness}(i), for the R\'enyi indices
    $\alpha=\tfrac12,1,2$, with $L=600$ sites and $512$ realizations for both
    models. Faint lines are single realizations. For the Hamiltonian both
    entropies grow as $\log\log t$: the dashed lines are fits
    $A_\alpha\log\log t+B_\alpha$ for $t\ge10^2$, with the fitted values at
    the top of each panel. For the circuit both entropies saturate.}
    \label{fig:quenched_disorder}
\end{figure*}

\subsection{Quenched disorder}
\label{sec:quenched}

In this section we sample the It\^o increments and the gates independently in
space at the first time step, and then we reapply them at all the following
time steps, as explained in Sec.~\ref{sec:types of randomness}(i), see also
Fig.~\ref{fig:models}(i).

The two models, the quantum circuit and the disordered Hamiltonian, behave in
opposite ways:
\begin{equation}
\begin{cases}
    S_A(t) = \mathcal{O}(\log \log t) \qquad &\text{for the Hamiltonian,} \\
    S_A(t) = \mathcal{O}(1) \qquad &\text{for the circuit.}
\end{cases}
\label{eq:quenched_disorder_entropy}
\end{equation}
The entropies of Eq.~\eqref{eq:quenched_disorder_entropy} has to be interpreted as the the behavior of both the Jordan--Wigner string and the time evolution operator. The OSEE of the
time evolution operator behaves in the same way, Fig.~\ref{fig:quenched_disorder}.

A one--dimensional free--fermion chain with disorder in space is expected to be
Anderson localized~\cite{Anderson_1958}, so that the entropy saturates. This is
what happens for the circuit~\cite{Ribeirp2021}. We claim that the Hamiltonian
escapes localization because of its symmetry class, in the sense of
Ref.~\cite{Atland_Zirnbaurer}.

With the noise frozen, $dW^t_j = W_j\,dt$, the Hamiltonian is a hopping
Hamiltonian with complex random couplings and no on--site term. The phases of
the couplings can be removed by the gauge transformation
$c_j\to e^{i\theta_j}c_j$, which leaves the entanglement unchanged. After this
transformation the quenched QSSEP Hamiltonian reads
\begin{equation}
\label{eq:quenched_hamiltonian}
    H_{\text{quenched}} = \sum_j |W_j| \left(c^\dagger_j c_{j+1} +c^\dagger_{j+1} c_{j} \right).
\end{equation}

The key difference between the Hamiltonian and the circuit is that the former
has a chiral symmetry, while the latter does not. Indeed, ordering the
single--particle basis into even sites $\{c^\dagger_{2j}\ket{0}\}$ and odd
sites $\{c^\dagger_{2j-1}\ket{0}\}$, the Hamiltonian~\eqref{eq:quenched_hamiltonian}
takes the block off--diagonal form
\begin{equation}
    H =
    \begin{pmatrix}
        0 & \mathcal{W} \\
        \mathcal{W}^T & 0
    \end{pmatrix},
    \label{eq:chiral_symmetry}
\end{equation}
where the bidiagonal matrix $\mathcal{W}$ contains the hopping amplitudes
$|W_j|$. A Hamiltonian that can be written as \eqref{eq:chiral_symmetry} has chiral
symmetry~\cite{Evers_2008}. In the circuit, instead, the Haar gates contain on-site terms, which break \eqref{eq:chiral_symmetry}.
For random hopping amplitudes with the same distribution on all the bonds, as
here, the chiral symmetry keeps the centre of the band
critical~\cite{Dyson_1953,Evers_2008}. In the language of spin chains,
Eq.~\eqref{eq:quenched_hamiltonian} is the random XX chain, which flows to an
infinite--randomness critical point~\cite{Fisher_1994}. At such critical points
the entanglement entropy after a quench grows as $\log\log t$. This was shown
for the random transverse--field Ising chain~\cite{Fischer_TFIM} in
Ref.~\cite{IgloiSzatmariLin_2012}, which also argues that the same ultraslow
growth holds for the random XY chain~\cite{IgloiJuhasz_2008}.

The comparison between the behavior of Eq.~\eqref{eq:quenched_disorder_entropy}
and the numerics is shown in Fig.~\ref{fig:quenched_disorder}; the agreement is good.

\subsection{Purely temporal disorder}

\begin{figure*}[!htb]
    \centering
    \includegraphics[width=0.88\textwidth]{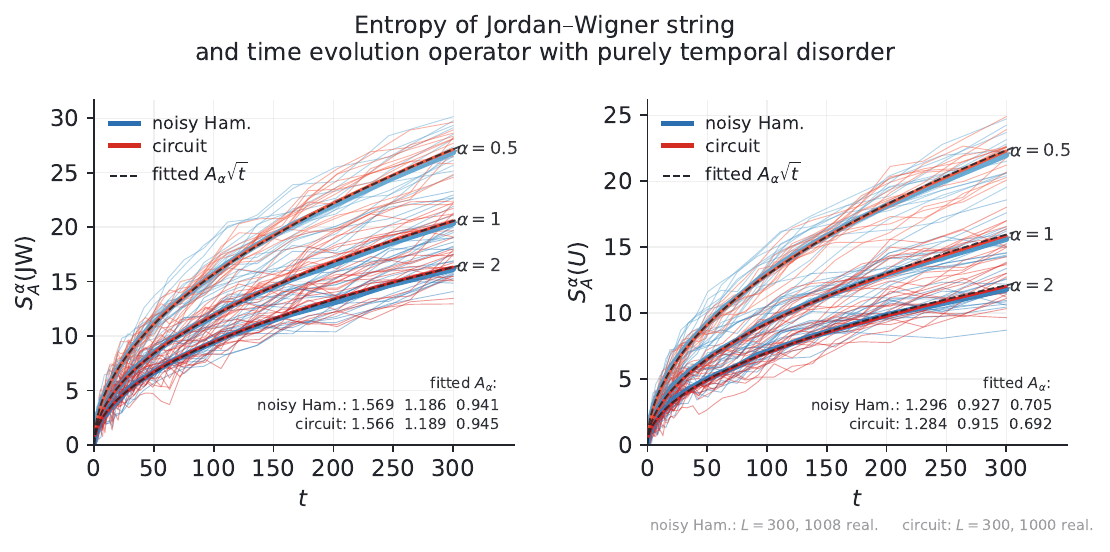}
    \caption{OSEE of the Jordan--Wigner string and the time evolution operator, evolved with purely temporal disorder as explained in Sec.~\ref{sec:types of randomness} and in Fig.~\ref{fig:models}. We consider the Renyi indices $\alpha=\tfrac12,1,2$. Faint lines are single realizations. We don't have a closed analytical expression for the entropies; the values of the $\sqrt t$ fit are shown in the bottom of the plots. We have taken $300$ sites and $1000$ realizations for the noisy Hamiltonian and $1008$ for the circuit. We observe very good agreement between the two models.  \emph{Left:} time evolution of the OSEE of the Jordan--Wigner string. \emph{Right:} time evolution of the OSEE of the time evolution operator.}
    \label{fig:purely_temporal_disorder}
\end{figure*}

Here we take the It\^o increments and the gates to be independent in time, but the same on all the odd bonds and the same on all the even bonds for each time step, as explained in Sec.~\ref{sec:types of randomness}(ii) and in Fig.~\ref{fig:models}(ii). The case $\nu=1$, that corresponds to a disorder that is translationally invariant by just one site, is special and is considered separately in App.~\ref{app:nu=1}. Otherwise, as long as $\nu \geq 2$, we find that the results are qualitatively the same for the random Hamiltonian and for the circuit.

In this case we obtain a \emph{square root growth of OSEE} of both the Jordan--Wigner string and the time evolution operator. The numerical result is shown in Fig.~\ref{fig:purely_temporal_disorder}.
Both the noisy Hamiltonian and the quantum circuit are \emph{not self-averaging}. We briefly discuss this numerically in App.~\ref{app:nu=2_selfaveraging}. We have taken $L=300$ sites and $1000$ realizations for the noisy Hamiltonian and $1008$ for the circuit.

\subsection{Fully random disorder}
\label{sec:fullyrandom}
In this Subsection, we take the It\^o increments and the gates to be independent in both space and time as explained in Sec.~\ref{sec:types of randomness}(iii), see also Fig.~\ref{fig:models}(iii).

As highlighted in the introduction, we obtain a \emph{square root growth of OSEE} of both the Jordan--Wigner string and the time evolution operator, see Fig.~\ref{fig:osee_full}. 
We have taken $L=300$ sites and $1008$ realizations for the Hamiltonian and $1000$ realizations for the circuit. 

In Sec.~\ref{sec:analytic}, we will compute analytically the prefactors of the OSEE of both the Jordan--Wigner string and the time evolution for the noisy Hamiltonian and the quantum circuit. Here we compare our numerics to these analytical results. The OSEE of the Jordan--Wigner string is predicted to grow as $S^\alpha_A(\rm{JW}) = 2s_\alpha \sqrt t +o(\sqrt t) $ where $s_\alpha$ is explicitly given in Eqs.~\eqref{eq:Renyi}-\eqref{eq:vN}. For the time-evolution operator the prediction is $S^\alpha_A(U) = c_\alpha \sqrt t + o(\sqrt t )$ where $c_\alpha$ is given by Eq.~\eqref{eq:c_alpha}. 

The numerical comparison is summarized in the following table,
$$
\begin{tabular}{c|ccc|ccc}
\hline\hline
 & \multicolumn{3}{c|}{$\mathrm{JW}$} & \multicolumn{3}{c}{$U$}\\
$\alpha$ & $2s_\alpha$ & circuit & Ham. & $c_\alpha$ & circuit & Ham.\\
\hline
$\tfrac12$ & $1.3938$ & $1.3940$ & $1.3856$ & $1.2210$ & $1.2198$ & $1.2109$\\
$1$        & $1.0475$ & $1.0475$ & $1.0404$ & $0.8998$ & $0.8991$ & $0.8925$\\
$2$        & $0.8283$ & $0.8282$ & $0.8231$ & $0.7051$ & $0.7044$ & $0.6992$\\
\hline\hline
\end{tabular}
$$
The first column gives the predicted analytical values, and the second and third columns give the fitted value extracted from the numerical results in the random circuit and in the random Hamiltonian. We fit the average over realizations with $a\sqrt t$ starting from times $t \geq 0.25 t_{\rm{max}}$ where $t_{\rm{max}} = 300$ in order to minimize short-time effects. The agreement is excellent.

\section{Analytic derivation of the \texorpdfstring{$\sqrt{t}$}{sqrt(t)} growth with fully random disorder}
\label{sec:analytic}

In this Section we show how to analytically compute the average OSEE plotted in Fig.~\ref{fig:osee_full} for any $\alpha>0$. The treatment below is valid for both models (Brownian Hamiltonian and random matchgate circuit) in the presence of disorder both in space and in time (case (iii) in Subsection~\ref{sec:types of randomness} and Fig.~\ref{fig:models}). For the other two types of disorder considered in Fig.~\ref{fig:models} we do not know an exact analytical treatment, and computing the OSEE remains an interesting open problem. 

Following the results of Section~\ref{sec:jw}, we need to compute the spectrum $\{\tau_k\}$ of the reduced correlation matrix $G^U_{BB}$ in Eq.~\eqref{eq:twofunctions}. It turns out that it is possible to compute the moments $m_n:= {\rm tr} [(G^U_{BB})^n] = \sum_k (\tau_k)^n$ under the assumption that the the Full Counting Statistics (FCS) defined below in Eq.~\eqref{def:FCS} is \emph{self--averaging}. This is enough to compute the asymptotic behavior of the OSEE at late times. We are not able to analytically prove the self-averaging property of the FCS, but we provide extensive numerical evidence for the validity of this assumption in Subsection~\ref{subsection:self_averaging}.

\subsection{Derivation of analytical results for the OSEE assuming self-averaging of the full counting statistics}

The calculation consists in three main steps. 
\begin{enumerate}[(1)]
\item One defines the Full Counting Statistics (FCS) of the particle number in the region $B=\{1,\dots L/2\}$ as
\begin{equation}
    \chi(\lambda):=\big\langle e^{\lambda N_B}\big\rangle.
    \label{def:FCS}
\end{equation}
All the cumulants of the particle number $N_B$ can be computed by deriving the logarithm of Eq.~\eqref{def:FCS} with respect to $\lambda$ and setting $\lambda =0$.

In a free fermionic system the FCS is a determinant for Gaussian states. This was first observed by Levitov in \cite{levitov_1993} and later generalized by Klich \cite{klich_2002}; for us the useful formula is 
\begin{equation}
\begin{aligned}
    \chi(\lambda) = \det(1+\omega\, G_B),
\end{aligned}
    \label{eq:Klich Levitov}
\end{equation}
with $\omega:=e^\lambda-1$.
The formula above is known as the Levitov formula and it is useful because it allows to relate the entanglement entropy to the full counting statistics \cite{klich_levitov_2009, song_2012}.
Taking the logarithm of the Levitov formula, we obtain
\begin{equation}
    \log\chi (\lambda)=\sum_{n\ge1}\frac{(-1)^{n-1}}{n}\Tr(G_B^n)\,\omega^n.
    \label{eq:FCS}
\end{equation}

\item We use the following relation between the QSSEP and the \emph{classical} symmetric simple exclusion process (SSEP)\cite{TonyJin_2019_openQSSEP}
\begin{equation}
    \mathbb{E}\big[\Tr(\rho\, e^{\sum_i a_i \hat n_i})\big] = \mathbb{E}\big[ e^{\sum_j a_j n_j}\big]_{\text{SSEP}}.
    \label{eq:QSSEP=SSEP}
\end{equation}
This equation implies that, on average, all the expectation values of observables diagonal in the occupation basis can be computed from the classical SSEP.

We claim that the identity \eqref{eq:QSSEP=SSEP} also holds for our Haar-random matchgate circuit. While we have not provided an analytical proof, we observe that all the results obtained from the arguments below are numerically consistent. For example, the traces of the power of the correlation matrix given in Eq.~\eqref{eq:moments} below are correctly reproduced by numerics in the circuit to all order we checked, see Fig.~\ref{fig:traces_u1}. 

\begin{figure}[!htb]
    \centering
    \includegraphics[width=\columnwidth]{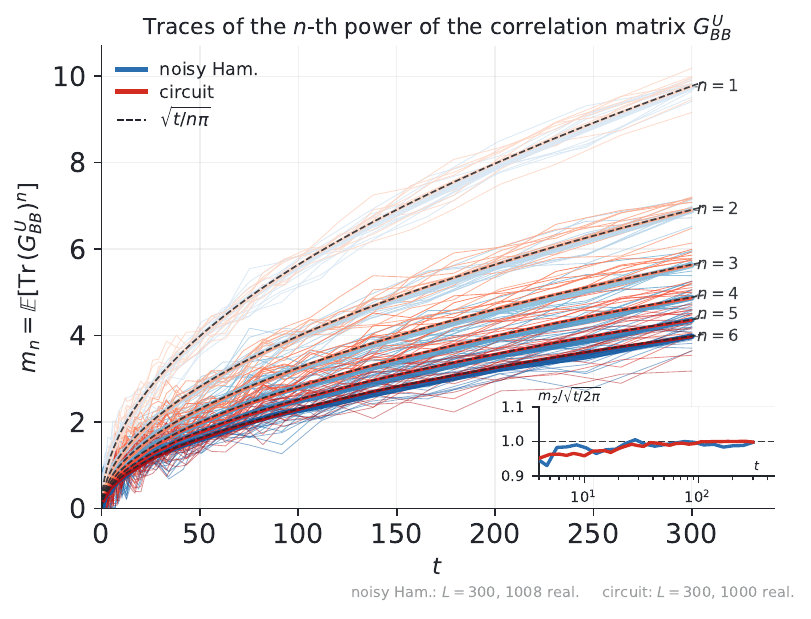}
    \caption{Trace moments $m_n=\mathbb{E}[\Tr (G^{U}_{BB})^n]$, $n=1,\dots,6$, for the
    two fully random models, $L=300$ sites, with $1008$ realizations for the noisy
    Hamiltonian and $1000$ for the circuit; faint lines are single realizations. \emph{Main panel:} moments $m_n$ against the closed form~\eqref{eq:moments} (dashed). \emph{Inset:} $m_2$ divided by its prediction, on a logarithmic time axis; it approaches $1$ for both models.}
    \label{fig:traces_u1}
\end{figure}

\begin{figure*}[!htb]
    \centering
    \includegraphics[width=0.85\textwidth]{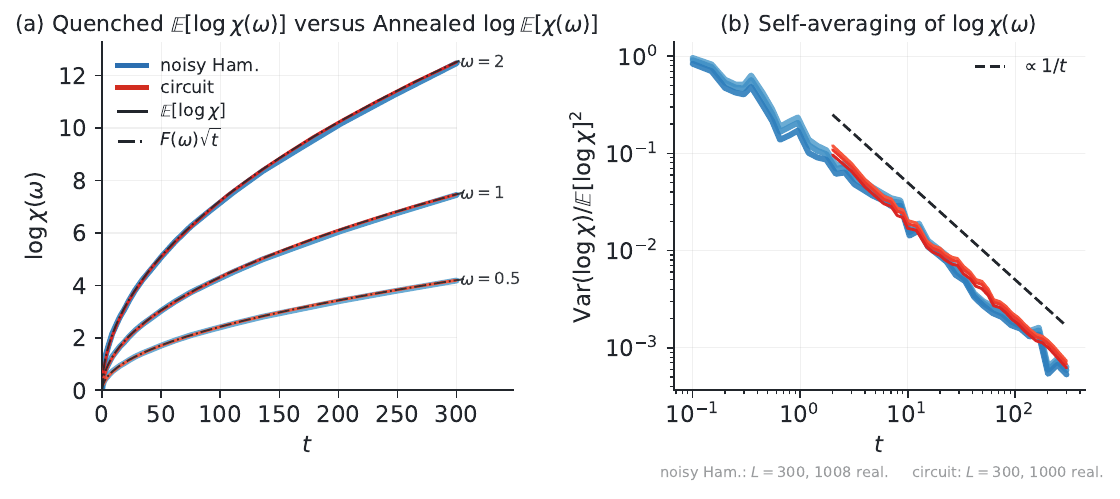}
    \caption{We show the simulation in favour of the self--averaging of the FCS \eqref{eq:self_averaing_FCS}. Both models are represented. (a) $\mathbb{E}[\log\chi(\lambda)]$ lies on the parameter--free SSEP line $F(\lambda)\sqrt{t}$ of Eq.~\eqref{eq:SSEP_F}. The SSEP LDF coincide with the annealed value of the FCS \eqref{eq:FCS=LDF}, hence the annealed and the quenched FCS overlaps. (b) The relative fluctuation of the FCS ${\rm Var}(\log\chi)/\mathbb{E}[\log\chi]^2$ vanishes as $1/t$.}
    \label{fig:selfav}
\end{figure*}

The SSEP is a one-dimensional classical model of particles hopping stochastically to nearest-neighbour sites \cite{Derrida_lectures_notes}. The hopping rate is symmetric: the jump rate to the left is the same as the one to the right. The only constraint in the SSEP is that two particles cannot be at the same site.
We introduce the integrated current $Q^t$, which measures the number of particles that have moved from region $A$ to region $B$ at time $t$.
The SSEP integrated current $Q^t$ through the cut \eqref{eq:bipatition}, satisfies large--deviation principle 
\begin{equation}
    \big\langle e^{\lambda Q^t}\big\rangle \asymp e^{\sqrt{t}\, F(\lambda)}
    \label{eq:LDF}
\end{equation}
The symbol $\asymp$ means that the ratio of the logarithm of both sites of Eq.~\eqref{eq:LDF} goes to $1$ at long times.
In particular, for a domain wall, the large deviation function $F(\omega)$ was derived by Derrida and Gerschenfeld \cite{Derrida_2009}
\begin{equation}
    F(\lambda)=\frac{1}{\sqrt{\pi}}\sum_{n\ge1}\frac{(-1)^{n+1}}{n^{3/2}}\,\omega^n,
    \label{eq:SSEP_F}
\end{equation}
with $\omega := e^\lambda -1$ as in Eq.~\eqref{eq:Klich Levitov}.
Using Eq.~\eqref{eq:QSSEP=SSEP} together with Eq.~\eqref{eq:SSEP_F}, we have
\begin{equation}
    \mathbb{E}\left[\langle e^{\lambda N_B} \rangle \right] = \mathbb{E}_{\text{SSEP}}\left[e^{\lambda Q^t} \right],
    \label{eq:FCS=LDF}
\end{equation}
which fixes the \emph{annealed average} $\log\mathbb{E}[\chi]$. Now we assume that the FCS is self--averaging,
\begin{equation}
    \log\mathbb{E}[\chi] = \mathbb{E}\left[\log \chi\right] \, + o(\sqrt t).
    \label{eq:self_averaing_FCS}
\end{equation}
Using the relation \eqref{eq:FCS}, together with the explicit expressions of Eq.~\eqref{eq:FCS=LDF} and Eq.~\eqref{eq:SSEP_F}, we obtain that, under the assumption \eqref{eq:self_averaing_FCS}, the traces of any power $n \geq 1$ of the correlation matrix are, on average, given by
\begin{equation}
    \mathbb{E}\left[\Tr (G^{U(t)}_{BB})^n\right] = \sqrt{\frac{t}{n\pi}} + o(\sqrt{t}).
    \label{eq:moments}
\end{equation}
Fig.~\ref{fig:traces_u1} shows the numerical simulation of the six moments $n=1,\dots,6$ against Eq.~\eqref{eq:moments}. We observe excellent agreement for both models, i.e. the noisy Hamiltonian and the quantum circuit.

\item We use the fact the moments \eqref{eq:moments} fix the Rényi entropy for any $\alpha > 0$. The reason is that for any analytic function $f$ with $f(0)=0$, we can Taylor-expand around $0$ to get $\Tr f(G^{U}_{BB})=\sum_n f_n\Tr (G^{U}_{BB})^n$; then inserting 
\begin{equation*}
    \frac{1}{\sqrt n}=\frac{1}{\sqrt \pi}\!\int_0^\infty \frac{ds}{\sqrt s}e^{-ns}
\end{equation*}
into Eq.~\eqref{eq:moments} and resumming over $n$ gives the formula
\begin{equation}
    \mathbb{E}[\Tr f(G_B)]=\frac{\sqrt{t}}{\pi}\int_0^\infty \frac{ds}{\sqrt{s}}\, f(e^{-s}) + o(\sqrt t).
    \label{eq:master}
\end{equation}
We apply this trick to the function $h_\alpha$ of Eq.~\eqref{eq:peschel} to get the OSEE of the Jordan-Wigner string. Keeping track of the factor two of Eq.~\eqref{eq:factor2}, this gives
\begin{equation}
\begin{aligned}
    &\mathbb{E}\big[S^\alpha_A\big(\mathrm{JW}(t)\big)\big]=2 s_\alpha\sqrt{t} + o(\sqrt{t}),\\
    &2s_\alpha=\frac{2}{\pi(1-\alpha)}\int_0^\infty\frac{ds}{\sqrt{s}}\ln\!\big[e^{-\alpha s}+(1-e^{-s})^\alpha\big].
\end{aligned}
    \label{eq:Renyi}
\end{equation}
The von Neumann limit $\alpha\to1$ gives
\begin{equation}
\begin{aligned}
    2 s_1 =  \frac{2}{\pi}\int_0^\infty\frac{ds}{\sqrt{s}} \Big[ s e^{-s} - (1-e^{-s})\ln(1-e^{-s}) \Big] \, .
\end{aligned}
    \label{eq:vN}
\end{equation}
The same trick can be applied to the function $f_\alpha$~\eqref{eq:falpha} to get the OSEE of the evolution operator 
\begin{equation}
\begin{aligned}
    \mathbb{E}\big[S^\alpha_A\big(U(t)\big)\big]&=c_\alpha\sqrt{t} + o(\sqrt t),\\
    c_\alpha&=\frac{2}{\pi}\int_0^\infty\frac{ds}{\sqrt{s}}\,
    h_\alpha\!\Big(\frac{1+\sqrt{1-e^{-s}}}{2}\Big) .
\end{aligned}
    \label{eq:c_alpha}
\end{equation}
\end{enumerate}
Eqs.\eqref{eq:Renyi}-\eqref{eq:c_alpha} are our main analytical results for the Jordan-Wigner string and the evolution operator. They are shown as dashed lines in Fig.~\ref{fig:osee_full}.

\subsection{Numerical evidence for the self--averaging property of the full counting statistics}
\label{subsection:self_averaging}
In Fig.~\ref{fig:selfav} we show numerical evidence that $\log \chi(\omega)$ is self-averaging.
In Fig.~\ref{fig:selfav}(a), the \emph{quenched average} of the logarithm of the FCS~\eqref{def:FCS}, $\mathbb{E}[\log \chi]$, agrees with the \emph{annealed average}, $\log \mathbb{E}[\chi]$, which in turn coincides with the large deviation function $\sqrt{t} F(\lambda)$ of the SSEP~\eqref{eq:FCS=LDF} at leading order in $\sqrt t$.

We can also relate the annealed and quenched averages via the cumulant expansion of $\log \mathbb{E}[\chi] = \log \mathbb{E}[e^{\log\chi}]$, the first two terms read
\begin{equation}
    \log \mathbb{E} [\chi] = \mathbb{E} [\log \chi] + \frac{1}{2} \left( \mathbb{E}\left[(\log\chi)^2\right] - \mathbb{E}[\log\chi]^2 \right)+\cdots.
\end{equation}
Then one needs to check that the second term is negligible. 
Numerically, we find that this is indeed the case. As shown in Fig.~\ref{fig:selfav}(b), the relative variance decays as
\begin{equation}
    \frac{\mathrm{Var}(\log \chi)}{\mathbb{E}[\log \chi]^2} := \frac{\mathbb{E}\left[(\log\chi)^2\right] - \mathbb{E}[\log\chi]^2}{\mathbb{E}[\log\chi]^2} \sim \frac{1}{t}.
    \label{eq:FCS_self_averaging_vanishing_relative_fluctuations}
\end{equation}
This allows us to conclude that
\begin{equation}
    \mathbb{E}[\log\chi] = \log\mathbb{E}[\chi] + o(\sqrt t)
    \label{eq:FCS_self_averaging_up_to_O(1)}
\end{equation}
as claimed in Eq.~\eqref{eq:self_averaing_FCS}.

\section{Discussion and conclusion}
\label{sec:conclusion}

\emph{Summary.}--- In this work, we have revisited the problem of the growth of the OSEE of local operators and of the evolution operator in quantum chains that can be mapped to free fermions. In contrast to previous works which focused on clean free fermion systems with translation-invariant and time-independent Hamiltonians~\cite{prosen2007operator,Prosen_OSEE_XY,Dubail_2017}, here we investigated models with randomness either in space or in time, or both. Our main results can be summarized as follows for the different cases:
\begin{enumerate}[(i)]
    \item \emph{Quenched disorder}: We found that in matchgate circuits the entropy of both the time evolution operator and the Jordan--Wigner string saturates, while for the disordered Hamiltonian the OSEE grows as $\log \log t$ for both operators. The results for this type of disorder are numerical.
    \item \emph{Purely temporal disorder}: The OSEE of the Jordan--Wigner string and the OSEE of the time evolution operator both grow \emph{diffusively}, i.e. $S^\alpha_A \sim \sqrt{t}$ for any Rényi index $\alpha >0$. For this type of disorder our results are numerical. 
    \item \emph{Fully random disorder}: Also in this case, the OSEE of the Jordan--Wigner string and the OSEE of the time evolution operator grow as $\mathcal O(\sqrt{t})$. For this type of disordered models our analysis relies on an interplay between numerical and analytical results. We have numerically validated the hypothesis that the full counting statistics is self-averaging. Relying on this assumption, we obtained closed formulas for OSEE of the Jordan-Wigner string and the evolution operator \eqref{eq:Renyi}-\eqref{eq:c_alpha}.
\end{enumerate}

\emph{Discussion.}--- Our most striking result concerns the OSEE of the Jordan-Wigner string in cases (ii) and (iii) above, which governs the growth of OSEE of all local spin operators involving a string. We found that it generically grows as $\mathcal O(\sqrt{t})$, in contrast with the $\mathcal O(\log t)$ growth previously found for the clean case~\cite{prosen2007operator,Prosen_OSEE_XY,Dubail_2017}. 

This $\mathcal O(\sqrt{t})$ growth is at odds with the widespread belief that the OSEE of local operators in systems with integrable dynamics should grow at most logarithmically~\cite{znidaric2008complexity,Alba_2019,Bertini_2020_II,pozsgay2026growth, alba2021diffusion,dowling2023scrambling}, at least if one regards these models of random free-fermion Hamiltonians and random matchgate circuits as `integrable'. They are clearly not `integrable' in the sense of Yang-Baxter integrability, but they are non-interacting and therefore their `complexity' is greatly reduced. It turns out this reduction of complexity is only partially reflected by the OSEE of local operators: the $\mathcal O(\sqrt{t})$ growth is slower than in chaotic quantum circuits where it goes as $\mathcal O(t)$~\cite{Nahum_2018,Bertini_2020}, but it is definitely much faster than logarithmic. This result should have consequences on the status of the OSEE of local operators in Heisenberg picture as a quantifier of the `complexity' of dynamics: it might not be such a great quantifier after all.

\emph{Future directions.}--- This work could be extended in a few directions. First, here we focused exclusively on free fermion models with conserved particle number, but it would be natural to consider also models without $U(1)$ conservation. The important question in that case is whether or not the dynamics in operator space still maps to an exclusion process similat to the one of Section~\ref{sec:analytic}; we plan to investigate this in subsequent work.

Second, it would be instructive to analytically compute the entanglement entropy of the domain wall, and hence the OSEE of the Jordan--Wigner string and the time evolution operator, for the purely temporal disorder. In this case, however, we don't have a classical underlying models as the SSEP; furthermore the full counting statistics is not self-averaging as explained in App.~\ref{app:nu=2_selfaveraging}.

Third, it might be interesting trying to develop a framework to analytically compute $\text{Var}[\log \chi(\lambda)] = o(\sqrt t)$ in Eq.~\eqref{eq:FCS_self_averaging_up_to_O(1)}. In fact, all the quantities here are "classical", because, at first order, they are fixed by $\log \mathbb{E}[\chi]$, which is determined by the classical SSEP \eqref{eq:FCS=LDF}. Thus, $\text{Var}[\log \chi(\lambda)]$ represent the quantum corrections to the quantities presented in this work, such as the entropies, the moments of the correlation matrix but also the cumulants of the particle number. 

Finally, it would be very interesting to go beyond free fermions and to ask if the diffusion behavior of the OSEE of local operators survives weak interactions, for instance in the models considered in Ref.~\cite{Nahum_2025_random_majorana}.

 \vspace{0.3cm}
{\bf Note added.} While we were finalizing this paper, a   preprint by Pozsgay appeared on the arXiv~\cite{pozsgay2026square}. Pozsgay reports the AI-assisted construction of a new model with an $\mathcal{O}(\sqrt{t})$ growth of the OSEE of local operators. 
Both Pozsgay's work and our own work are motivated by the search for counterexamples to the belief that the OSEE of local operators grows as $\mathcal{O}(\log t)$ under integrable dynamics. We both identify models where the OSEE instead grows as $\mathcal{O}(\sqrt{t})$. Yet, as far as we can see there is no relation between our models. Pozsgay's model is a quantum circuit model with no randomness that falls in the class of Yang-Baxter-integrable systems, while the models we have investigated here are random free fermion models not related to Yang-Baxter integrability.

\begin{figure*}[!htb]
    \centering
    \includegraphics[width=0.85\textwidth]{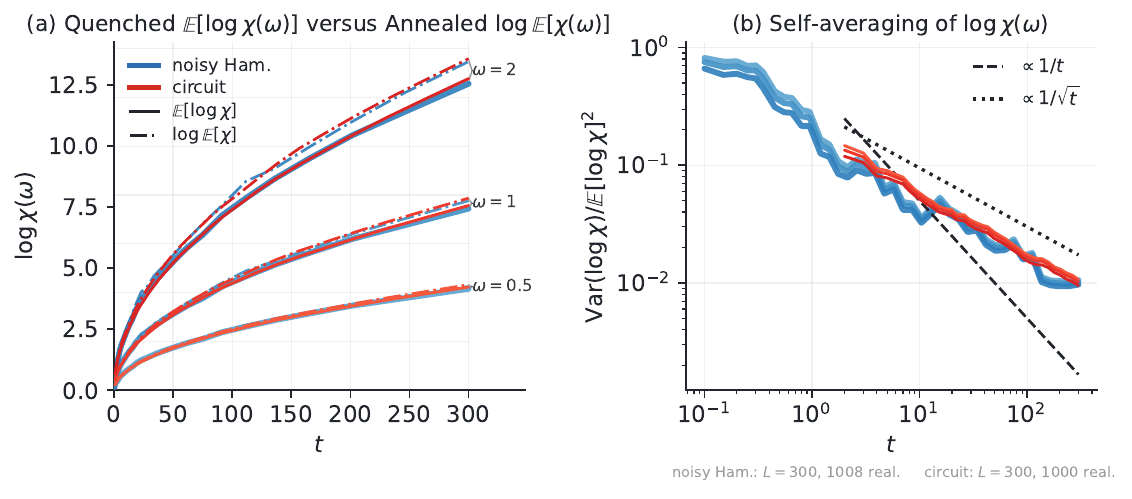}
    \caption{We show the numerical simulation of the FCS \eqref{def:FCS} and the relative fluctuations \eqref{eq:FCS_self_averaging_vanishing_relative_fluctuations}. Both models are represented. (a) $\mathbb{E}[\log\chi(\omega)]$ and $\log \mathbb{E}[\chi(\omega)]$ are plotted, the two quantities do not overlap, in stark contrast with the fully random case \ref{fig:selfav}. (b) The relative fluctuation of the FCS ${\rm Var}(\log\chi)/\mathbb{E}[\log\chi]^2$ vanishes as $1/\sqrt{t}$, so $\text{Var}[\log\chi]=\mathcal{O}(\sqrt{t})$.}
    \label{fig:selfav_nu=2}
\end{figure*}

\begin{acknowledgments}
    We thank Bruno Bertini, João Costa, Igor Gornyi, Sasha Mirlin, Lorenzo Piroli and Igor Poboiko for their interest in this project and for useful discussions.

This work of the Interdisciplinary Thematic Institute QMat, as part of the ITI 2021-2028 program of the University of Strasbourg, CNRS and Inserm, was supported by IdEx Unistra (ANR 10 IDEX 0002), and by SFRI STRAT’US project (ANR 20 SFRI 0012) and EUR QMAT ANR-17-EURE-0024 under the framework of the French Investments for the Future Program. This work also benefited from support managed by the Agence Nationale de la Recherche under the France 2030 program  (ANR-22-CMAS-0001), and under the ANR-PRME project `UNIOPEN' (ANR-22-CE30-0004-01). Computations were carried out using resources of the High Performance Computing Center of the University of Strasbourg, funded by Equip@Meso (as part of the Investments for the Future Program) and CPER Alsacalcul/Big Data.

\end{acknowledgments}

\appendix

\section{The FCS with purely temporal disorder is not self-averaging.}
\label{app:nu=2_selfaveraging}

In this appendix we numerically analyze the self averaging property of the Full Counting Statistics (FCS) for purely temporal disorder, namely the type of disorder (ii) in Fig.~\ref{fig:models}. 

We consider the self-averaging of the FCS, defined in Eq.~\eqref{def:FCS}. We observe that this quantity is the central object to study: in fact, it is directly related to the entanglement \cite{klich_levitov_2009, song_2012}, thus if the FCS is not self-averaging we expect the entropy not to be self-averaging either. 

The quantities considered are the same of Fig.~\ref{fig:selfav}, in particular the relative fluctuations is defined in  Eq.~\eqref{eq:FCS_self_averaging_vanishing_relative_fluctuations}. 

The numerical result is shown in Fig.~\ref{fig:selfav_nu=2}. The result is completely different from the one observed in the fully random case \ref{fig:selfav}. The quenched $\mathbb{E}[\log\chi(\omega)]$ and annealed averages $\log \mathbb{E}[\chi(\omega)]$ do not overlap and the relative fluctuations decay as $1/\sqrt t$ as opposed to $1/t$ in the fully random case \eqref{eq:FCS_self_averaging_vanishing_relative_fluctuations}.

\section{\texorpdfstring{${\nu=1}$}{ν=1} is non--generic.}
\label{app:nu=1}

\begin{figure}[!htb]
    \centering
    \includegraphics[width=\columnwidth]{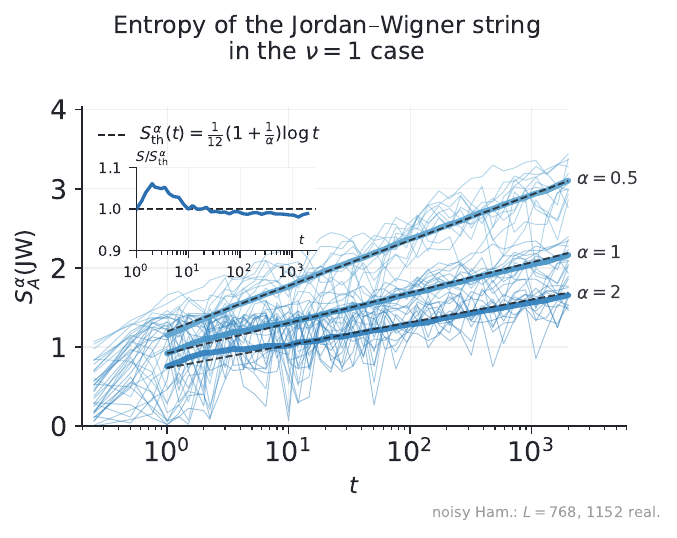}
    \caption{OSEE of the Jordan--Wigner string for temporal disorder and translational invariance of $\nu=1$ sites. The dashed line is the theoretical prediction given in Eq.~\eqref{eq:nu=1_entnaglement}. Faint lines are single realizations. We show three values of the Renyi index $\alpha=0.5,1,2$. The inset represents the fraction ${S^1}/{S^1_{\rm{th}}}$ in log scale, which approaches $1$ at large times. We have taken $L=768$ sites and $1152$ realizations. Contrary to all the other cases with temporal disorder analyzed in this work, here the OSEE of the Jordan--Wigner string grows as $\log(t)$.}
    \label{fig:nu=1}
\end{figure}

This appendix shows why the QSSEP with translational symmetry by one site is non generic. 
For the $\nu=1$ QSSEP the OSEE of the Jordan--Wigner string grows logarithmically 
\begin{equation}
    S^\alpha_A = \frac{1}{12}\left(1+\frac{1}{\alpha} \right) \log t + \mathcal{O}(1).
    \label{eq:nu=1_entnaglement}
\end{equation}
The coefficient of the logarithm is half the one in Eq.~\eqref{eq:temporal_log} as already observed in \cite{Alba_nuQSSEP_GHD}. 

The results of Eq.~\eqref{eq:nu=1_entnaglement} are shown in Fig.~\ref{fig:nu=1}. They very good agreement with the theoretical formula.  

\emph{Why $\nu=1$ is special: coherent quasiparticles.}---
When a single It\^o increment is shared by all bonds the Hamiltonian is translation invariant \emph{at every instant}, so momentum is conserved in every realization. The single--particle problem is then diagonal in $k$ for all times: each mode keeps its momentum and acquires only a random phase, so that its group velocity $v_k(t)$ is a white noise. The quasiparticles therefore remain \emph{coherent}: they are the same excitations as in the clean chain, merely carried along a light cone whose radius $\int_0^t v_k(s)\,ds$ performs a random walk instead of growing linearly. The melting domain wall keeps the structure it has in the clean chain, and the conformal counting that gives Eq.~\eqref{eq:nu=1_entnaglement} survives with $t$ replaced by the fluctuating cone, whose radius grows as $\sqrt t$.

The coherence of the quasi-particles is lost whenever a momentum $k_1$ scatter with $k_2 \neq k_1$. So, spatial randomness ($\nu>1$) scatters between different $k$ and destroys the logarithm. We observe that, by the same argument, a translationally invariant system  by one site with temporal randomness but without $U(1)$ symmetry will break the logarithm, because momenta $k$ and $-k$ can scatter.

\bibliography{bib}

@article{Vestraete_2006,
  title = {Matrix product states represent ground states faithfully},
  author = {Verstraete, F. and Cirac, J. I.},
  journal = {Phys. Rev. B},
  volume = {73},
  issue = {9},
  pages = {094423},
  numpages = {8},
  year = {2006},
  month = {Mar},
  publisher = {American Physical Society},
  doi = {10.1103/PhysRevB.73.094423},
  url = {https://link.aps.org/doi/10.1103/PhysRevB.73.094423}
}

@article{Hastings_2006,
  title = {Solving gapped Hamiltonians locally},
  author = {Hastings, M. B.},
  journal = {Phys. Rev. B},
  volume = {73},
  issue = {8},
  pages = {085115},
  numpages = {13},
  year = {2006},
  month = {Feb},
  publisher = {American Physical Society},
  doi = {10.1103/PhysRevB.73.085115},
  url = {https://link.aps.org/doi/10.1103/PhysRevB.73.085115}
}

@article{Vestraete_2008,
  title = {Entropy Scaling and Simulability by Matrix Product States},
  author = {Schuch, Norbert and Wolf, Michael M. and Verstraete, Frank and Cirac, J. Ignacio},
  journal = {Phys. Rev. Lett.},
  volume = {100},
  issue = {3},
  pages = {030504},
  numpages = {4},
  year = {2008},
  month = {Jan},
  publisher = {American Physical Society},
  doi = {10.1103/PhysRevLett.100.030504},
  url = {https://link.aps.org/doi/10.1103/PhysRevLett.100.030504}
}

@article{Hartmann_2009,
  title = {Density Matrix Renormalization Group in the Heisenberg Picture},
  author = {Hartmann, Michael J. and Prior, Javier and Clark, Stephen R. and Plenio, Martin B.},
  journal = {Phys. Rev. Lett.},
  volume = {102},
  issue = {5},
  pages = {057202},
  numpages = {4},
  year = {2009},
  month = {Feb},
  publisher = {American Physical Society},
  doi = {10.1103/PhysRevLett.102.057202},
  url = {https://link.aps.org/doi/10.1103/PhysRevLett.102.057202}
}

@article{Cirac_2015,
  title = {Variational Matrix Product Operators for the Steady State of Dissipative Quantum Systems},
  author = {Cui, Jian and Cirac, J. Ignacio and Ba\~nuls, Mari Carmen},
  journal = {Phys. Rev. Lett.},
  volume = {114},
  issue = {22},
  pages = {220601},
  numpages = {5},
  year = {2015},
  month = {Jun},
  publisher = {American Physical Society},
  doi = {10.1103/PhysRevLett.114.220601},
  url = {https://link.aps.org/doi/10.1103/PhysRevLett.114.220601}
}

@article{dowling2023scrambling,
  title = {Scrambling Is Necessary but Not Sufficient for Chaos},
  author = {Dowling, Neil and Kos, Pavel and Modi, Kavan},
  journal = {Phys. Rev. Lett.},
  volume = {131},
  issue = {18},
  pages = {180403},
  numpages = {6},
  year = {2023},
  month = {Nov},
  publisher = {American Physical Society},
  doi = {10.1103/PhysRevLett.131.180403},
  url = {https://link.aps.org/doi/10.1103/PhysRevLett.131.180403}
}

@article{alba2025more,
doi = {10.1088/1751-8121/adc9e6},
url = {https://doi.org/10.1088/1751-8121/adc9e6},
year = {2025},
month = {apr},
publisher = {IOP Publishing},
volume = {58},
number = {17},
pages = {175003},
author = {Alba, Vincenzo},
title = {More on the operator space entanglement (OSE): Rényi OSE, revivals, and integrability breaking*},
journal = {Journal of Physics A: Mathematical and Theoretical},
}

@article{alba2021diffusion,
  title = {Diffusion and operator entanglement spreading},
  author = {Alba, Vincenzo},
  journal = {Phys. Rev. B},
  volume = {104},
  issue = {9},
  pages = {094410},
  numpages = {8},
  year = {2021},
  month = {Sep},
  publisher = {American Physical Society},
  doi = {10.1103/PhysRevB.104.094410},
  url = {https://link.aps.org/doi/10.1103/PhysRevB.104.094410}
}

@article{dowling2025bridging,
  title = {Bridging Entanglement and Magic Resources within Operator Space},
  author = {Dowling, Neil and Modi, Kavan and White, Gregory A. L.},
  journal = {Phys. Rev. Lett.},
  volume = {135},
  issue = {16},
  pages = {160201},
  numpages = {9},
  year = {2025},
  month = {Oct},
  publisher = {American Physical Society},
  doi = {10.1103/c7k1-xcwy},
  url = {https://link.aps.org/doi/10.1103/c7k1-xcwy}
}

@misc{dowling_2026_classicalsimulabilityoperatorentanglement,
      title={Classical Simulability from Operator Entanglement Scaling}, 
      author={Neil Dowling},
      year={2026},
      eprint={2603.05656},
      archivePrefix={arXiv},
      primaryClass={quant-ph},
      url={https://arxiv.org/abs/2603.05656}, 
}

@article{Noh2020efficientclassical,
  doi = {10.22331/q-2020-09-11-318},
  url = {https://doi.org/10.22331/q-2020-09-11-318},
  title = {Efficient classical simulation of noisy random quantum circuits in one dimension},
  author = {Noh, Kyungjoo and Jiang, Liang and Fefferman, Bill},
  journal = {{Quantum}},
  issn = {2521-327X},
  publisher = {{Verein zur F{\"{o}}rderung des Open Access Publizierens in den Quantenwissenschaften}},
  volume = {4},
  pages = {318},
  month = sep,
  year = {2020}
}

@article{preisser2023comparing,
  title = {Comparing bipartite entropy growth in open-system matrix-product simulation methods},
  author = {Preisser, Guillermo and Wellnitz, David and Botzung, Thomas and Schachenmayer, Johannes},
  journal = {Phys. Rev. A},
  volume = {108},
  issue = {1},
  pages = {012616},
  numpages = {15},
  year = {2023},
  month = {Jul},
  publisher = {American Physical Society},
  doi = {10.1103/PhysRevA.108.012616},
  url = {https://link.aps.org/doi/10.1103/PhysRevA.108.012616}
}

@article{Zanardi_2001,
  title = {Entanglement of quantum evolutions},
  author = {Zanardi, Paolo},
  journal = {Phys. Rev. A},
  volume = {63},
  issue = {4},
  pages = {040304(R)},
  numpages = {4},
  year = {2001},
  month = {Mar},
  publisher = {American Physical Society},
  doi = {10.1103/PhysRevA.63.040304},
  url = {https://link.aps.org/doi/10.1103/PhysRevA.63.040304}
}

@article{li2025highly,
  title = {Highly Entangled Stationary States from Strong Symmetries},
  author = {Li, Yahui and Pollmann, Frank and Read, Nicholas and Sala, Pablo},
  journal = {Phys. Rev. X},
  volume = {15},
  issue = {1},
  pages = {011068},
  numpages = {37},
  year = {2025},
  month = {Mar},
  publisher = {American Physical Society},
  doi = {10.1103/PhysRevX.15.011068},
  url = {https://link.aps.org/doi/10.1103/PhysRevX.15.011068}
}

@article{rath2023entanglement,
  title = {Entanglement Barrier and its Symmetry Resolution: Theory and Experimental Observation},
  author = {Rath, Aniket and Vitale, Vittorio and Murciano, Sara and Votto, Matteo and Dubail, J{\'e}r{\^o}me and Kueng, Richard and Branciard, Cyril and Calabrese, Pasquale and Vermersch, Beno{\^{\i}}t},
  journal = {PRX Quantum},
  volume = {4},
  issue = {1},
  pages = {010318},
  numpages = {39},
  year = {2023},
  month = {Feb},
  publisher = {American Physical Society},
  doi = {10.1103/PRXQuantum.4.010318},
  url = {https://link.aps.org/doi/10.1103/PRXQuantum.4.010318}
}

@article{zhang2025operator,
  title = {Operator entanglement in SU(2)-symmetric dissipative quantum many-body dynamics},
  author = {Zhang, Lin},
  journal = {Phys. Rev. B},
  volume = {112},
  issue = {14},
  pages = {144303},
  numpages = {10},
  year = {2025},
  month = {Oct},
  publisher = {American Physical Society},
  doi = {10.1103/d8kg-h1t7},
  url = {https://link.aps.org/doi/10.1103/d8kg-h1t7}
}

@article{Prosen_2007,
  title = {Is the efficiency of classical simulations of quantum dynamics related to integrability?},
  author = {Prosen, Tomazifmmode \check{z}\else \v{z}\fi{} and \ifmmode \check{Z}\else \v{Z}\fi{}nidari\ifmmode \check{c}\else \v{c}\fi{}, Marko},
  journal = {Phys. Rev. E},
  volume = {75},
  issue = {1},
  pages = {015202(R)},
  numpages = {4},
  year = {2007},
  month = {Jan},
  publisher = {American Physical Society},
  doi = {10.1103/PhysRevE.75.015202},
  url = {https://link.aps.org/doi/10.1103/PhysRevE.75.015202}
}

@article{Prosen_OSEE_XY,
  title = {Operator space entanglement entropy in $XY$ spin chains},
  author = {Pi{\ifmmode \check{z}\else \v{z}\fi{}}orn, Iztok and Prosen, Toma{\ifmmode \check{z}\else \v{z}\fi{}}},
  journal = {Phys. Rev. B},
  volume = {79},
  issue = {18},
  pages = {184416},
  numpages = {4},
  year = {2009},
  month = {May},
  publisher = {American Physical Society},
  doi = {10.1103/PhysRevB.79.184416},
  url = {https://link.aps.org/doi/10.1103/PhysRevB.79.184416}
}

@article{prosen2007operator,
  title = {Operator space entanglement entropy in a transverse Ising chain},
  author = {Prosen, Toma{\v{z}} and Pi{\v{z}}orn, Iztok},
  journal = {Phys. Rev. A},
  volume = {76},
  issue = {3},
  pages = {032316},
  numpages = {5},
  year = {2007},
  month = {Sep},
  publisher = {American Physical Society},
  doi = {10.1103/PhysRevA.76.032316},
  url = {https://link.aps.org/doi/10.1103/PhysRevA.76.032316}
}

@article{Dubail_2017,
   title={Entanglement scaling of operators: a conformal field theory approach, with a glimpse of simulability of long-time dynamics in 1 + 1d},
   volume={50},
   ISSN={1751-8121},
   url={http://dx.doi.org/10.1088/1751-8121/aa6f38},
   DOI={10.1088/1751-8121/aa6f38},
   number={23},
   journal={Journal of Physics A: Mathematical and Theoretical},
   publisher={IOP Publishing},
   author={Dubail, J},
   year={2017},
   month=May, pages={234001} }

@article{Zhou_OSEE_evolution_operator_chaotic_systems,
  title = {Operator entanglement entropy of the time evolution operator in chaotic systems},
  author = {Zhou, Tianci and Luitz, David J.},
  journal = {Phys. Rev. B},
  volume = {95},
  issue = {9},
  pages = {094206},
  numpages = {15},
  year = {2017},
  month = {Mar},
  publisher = {American Physical Society},
  doi = {10.1103/PhysRevB.95.094206},
  url = {https://link.aps.org/doi/10.1103/PhysRevB.95.094206}
}

@article{Alba_2019,
   title={Operator Entanglement in Interacting Integrable Quantum Systems: the Case of the Rule 54 Chain},
   volume={122},
   DOI={10.1103/PhysRevLett.122.250603},
   number={25},
   journal={Physical Review Letters},
   publisher={American Physical Society (APS)},
   author={Alba, Vincenzo and Dubail, J{\'e}r{\^o}me and Medenjak, Marko},
   year={2019},
   month=Jun, pages={250603} }

@article{znidaric2008complexity,
  title = {Complexity of thermal states in quantum spin chains},
  author = {{\ifmmode \check{Z}\else \v{Z}\fi{}}nidari{\ifmmode \check{c}\else \v{c}\fi{}}, Marko and Prosen, Toma{\ifmmode \check{z}\else \v{z}\fi{}} and Pi{\ifmmode \check{z}\else \v{z}\fi{}}orn, Iztok},
  journal = {Phys. Rev. A},
  volume = {78},
  issue = {2},
  pages = {022103},
  numpages = {5},
  year = {2008},
  month = {Aug},
  publisher = {American Physical Society},
  doi = {10.1103/PhysRevA.78.022103},
  url = {https://link.aps.org/doi/10.1103/PhysRevA.78.022103},
}

@misc{pozsgay2026square,
      title={Square-root growth of operator entanglement in an integrable brickwork circuit}, 
      author={Balázs Pozsgay},
      year={2026},
      eprint={2609.39401},
      archivePrefix={arXiv},
      primaryClass={cond-mat.stat-mech},
}

@article{medenjak2022operator,
doi = {10.1088/1751-8121/ac8fc4},
url = {https://doi.org/10.1088/1751-8121/ac8fc4},
year = {2022},
month = {sep},
publisher = {IOP Publishing},
volume = {55},
number = {40},
pages = {404002},
author = {Medenjak, Marko},
title = {Operator spreading in quantum hardcore gases},
journal = {Journal of Physics A: Mathematical and Theoretical},
}

@article{wellnitz2022rise,
  title={Rise and fall, and slow rise again, of operator entanglement under dephasing},
  author={Wellnitz, David and Preisser, Guillermo and Alba, Vincenzo and Dubail, Jerome and Schachenmayer, Johannes},
  journal={Physical Review Letters},
  volume={129},
  number={17},
  pages={170401},
  year={2022},
  publisher={APS}
}

@article{rakovszky2022dissipation,
  title = {Dissipation-assisted operator evolution method for capturing hydrodynamic transport},
  author = {Rakovszky, Tibor and von Keyserlingk, C. W. and Pollmann, Frank},
  journal = {Phys. Rev. B},
  volume = {105},
  issue = {7},
  pages = {075131},
  numpages = {10},
  year = {2022},
  month = {Feb},
  publisher = {American Physical Society},
  doi = {10.1103/PhysRevB.105.075131},
  url = {https://link.aps.org/doi/10.1103/PhysRevB.105.075131}
}

@article{Muth_2011_numerics,
  title = {Dynamical Simulation of Integrable and Nonintegrable Models in the Heisenberg Picture},
  author = {Muth, Dominik and Unanyan, Razmik G. and Fleischhauer, Michael},
  journal = {Phys. Rev. Lett.},
  volume = {106},
  issue = {7},
  pages = {077202},
  numpages = {4},
  year = {2011},
  month = {Feb},
  publisher = {American Physical Society},
  doi = {10.1103/PhysRevLett.106.077202},
  url = {https://link.aps.org/doi/10.1103/PhysRevLett.106.077202}
}

@article{Gopalakrishnan_2026,
	title = {Long-time limits of local operator entanglement in interacting integrable models},
	pages = {162},
	author = {Jacoby, J. Alexander and Gopalakrishnan, Sarang},
	journal = {SciPost Phys.},
	volume = {20},
	year = {2026},
	publisher = {SciPost},
	doi = {10.21468/SciPostPhys.20.6.162},
	url = {https://scipost.org/10.21468/SciPostPhys.20.6.162}
}

@misc{Nahum_2018,
      title={Coarse-grained dynamics of operator and state entanglement}, 
      author={Cheryne Jonay and David A. Huse and Adam Nahum},
      year={2018},
      eprint={1803.00089},
      archivePrefix={arXiv},
      primaryClass={cond-mat.stat-mech},
      url={https://arxiv.org/abs/1803.00089}, 
}

@misc{dowling_2026,
      title={Page Curve for Local-Operator Entanglement from Free Probability}, 
      author={Neil Dowling and Silvia Pappalardi},
      year={2026},
      eprint={2605.02995},
      archivePrefix={arXiv},
      primaryClass={quant-ph},
      url={https://arxiv.org/abs/2605.02995}, 
}

@misc{correr_2026,
      title={A derivation of the late-time volume law for local operator entanglement}, 
      author={Guilherme Ilario Correr and John Goold and Marco Cattaneo},
      year={2026},
      eprint={2603.25387},
      archivePrefix={arXiv},
      primaryClass={quant-ph},
      url={https://arxiv.org/abs/2603.25387}, 
}

@misc{pozsgay2026growth,
      title={On the growth of operator entanglement in brickwork circuits with Yang--Baxter gates}, 
      author={Balázs Pozsgay},
      year={2026},
      eprint={2609.05121},
      archivePrefix={arXiv},
      primaryClass={cond-mat.stat-mech},
      url={https://arxiv.org/abs/2609.05121}, 
}

@article{Bertini_2020,
   title={Operator Entanglement in Local Quantum Circuits I: Chaotic Dual-Unitary Circuits},
   volume={8},
   DOI={10.21468/SciPostPhys.8.4.067},
   number={4},
   journal={SciPost Physics},
   publisher={Stichting SciPost},
   author={Bertini, Bruno and Kos, Pavel and Prosen, Tomazv{z}},
   year={2020},
   month=Apr, pages={067} }

@article{Bertini_2020_II,
	title = {Operator Entanglement in Local Quantum Circuits II: Solitons in Chains of Qubits},
	pages = {068},
	author = {Bertini, Bruno and Kos, Pavel and Prosen, Tomaz},
	journal = {SciPost Phys.},
	volume = {8},
	year = {2020},
	publisher = {SciPost},
	doi = {10.21468/SciPostPhys.8.4.068},
	url = {https://scipost.org/10.21468/SciPostPhys.8.4.068}
}

@article{Murciano_2024,
   title={More on symmetry resolved operator entanglement},
   volume={57},
   ISSN={1751-8121},
   url={http://dx.doi.org/10.1088/1751-8121/ad30d1},
   DOI={10.1088/1751-8121/ad30d1},
   number={14},
   journal={Journal of Physics A: Mathematical and Theoretical},
   publisher={IOP Publishing},
   author={Murciano, Sara and Dubail, Jerome and Calabrese, Pasquale},
   year={2024},
   month=Mar, pages={145002} }

@misc{TanProsen_2026,
   title={Logarithmic growth of operator entanglement in a clean non-integrable circuit},
   author={Tan, Mao Tian and Prosen, Tomazv{z}},
   year={2026},
   eprint={2603.19363},
   archivePrefix={arXiv},
   primaryClass={cond-mat.stat-mech},
   url={https://arxiv.org/abs/2603.19363} }

@article{Anderson_1958,
   title={Absence of diffusion in certain random lattices},
   author={Anderson, P. W.},
   journal={Phys. Rev.},
   volume={109},
   pages={1492},
   year={1958},
   doi={10.1103/PhysRev.109.1492} }

@article{Valiant_2002,
   title={Quantum circuits that can be simulated classically in polynomial time},
   author={Valiant, Leslie G.},
   journal={SIAM J. Comput.},
   volume={31},
   pages={1229},
   year={2002},
   doi={10.1137/S0097539700377025} }

@article{TerhalDiVincenzo_2002,
   title={Classical simulation of noninteracting-fermion quantum circuits},
   author={Terhal, Barbara M. and DiVincenzo, David P.},
   journal={Phys. Rev. A},
   volume={65},
   pages={032325},
   year={2002},
   doi={10.1103/PhysRevA.65.032325} }

@misc{Knill_2001,
   title={Fermionic linear optics and matchgates},
   author={Knill, Emanuel},
   year={2001},
   eprint={quant-ph/0108033},
   archivePrefix={arXiv},
   primaryClass={quant-ph},
   url={https://arxiv.org/abs/quant-ph/0108033} }

@article{JozsaMiyake_2008,
   title={Matchgates and classical simulation of quantum circuits},
   author={Jozsa, Richard and Miyake, Akimasa},
   journal={Proc. R. Soc. A},
   volume={464},
   pages={3089},
   year={2008},
   doi={10.1098/rspa.2008.0189} }

@article{TonyJin_2019,
	title = {Equilibrium fluctuations in maximally noisy extended quantum systems},
	pages = {045},
	author = {Bauer, Michel and Bernard, Denis and Jin, Tony},
	journal = {SciPost Phys.},
	volume = {6},
	year = {2019},
	publisher = {SciPost},
	doi = {10.21468/SciPostPhys.6.4.045},
	url = {https://scipost.org/10.21468/SciPostPhys.6.4.045}
}

@article{maccormack2021operator,
  title = {Operator and entanglement growth in nonthermalizing systems: Many-body localization and the random singlet phase},
  author = {MacCormack, Ian and Tan, Mao Tian and Kudler-Flam, Jonah and Ryu, Shinsei},
  journal = {Phys. Rev. B},
  volume = {104},
  issue = {21},
  pages = {214202},
  numpages = {12},
  year = {2021},
  month = {Dec},
  publisher = {American Physical Society},
  doi = {10.1103/PhysRevB.104.214202},
  url = {https://link.aps.org/doi/10.1103/PhysRevB.104.214202}
}

@article{kudler2021entanglement,
  title = {Entanglement of local operators and the butterfly effect},
  author = {Kudler-Flam, Jonah and Nozaki, Masahiro and Ryu, Shinsei and Tan, Mao Tian},
  journal = {Phys. Rev. Res.},
  volume = {3},
  issue = {3},
  pages = {033182},
  numpages = {25},
  year = {2021},
  month = {Aug},
  publisher = {American Physical Society},
  doi = {10.1103/PhysRevResearch.3.033182},
  url = {https://link.aps.org/doi/10.1103/PhysRevResearch.3.033182}
}

@article{TonyJin_2019_openQSSEP,
  title = {Open Quantum Symmetric Simple Exclusion Process},
  author = {Bernard, Denis and Jin, Tony},
  journal = {Phys. Rev. Lett.},
  volume = {123},
  issue = {8},
  pages = {080601},
  numpages = {7},
  year = {2019},
  month = {Aug},
  publisher = {American Physical Society},
  doi = {10.1103/PhysRevLett.123.080601},
  url = {https://link.aps.org/doi/10.1103/PhysRevLett.123.080601}
}

@article{Piroli_2021_entanglementQSSEP,
  title = {Entanglement distribution in the quantum symmetric simple exclusion process},
  author = {Bernard, Denis and Piroli, Lorenzo},
  journal = {Phys. Rev. E},
  volume = {104},
  issue = {1},
  pages = {014146},
  numpages = {6},
  year = {2021},
  month = {Jul},
  publisher = {American Physical Society},
  doi = {10.1103/PhysRevE.104.014146},
  url = {https://link.aps.org/doi/10.1103/PhysRevE.104.014146}
}

@article{Essler_2022_QSSEP_useful_formulas,
	title = {Dynamics of fluctuations in quantum simple exclusion processes},
	pages = {042},
	author = {Bernard, Denis and Essler, Fabian H. L. and Hruza, Ludwig and Medenjak, Marko},
	journal = {SciPost Phys.},
	volume = {12},
	year = {2022},
	publisher = {SciPost},
	doi = {10.21468/SciPostPhys.12.1.042},
	url = {https://scipost.org/10.21468/SciPostPhys.12.1.042}
}

@article{Hruza_2023_QSSEP_free_probability,
  title = {Coherent Fluctuations in Noisy Mesoscopic Systems, the Open Quantum SSEP, and Free Probability},
  author = {Hruza, Ludwig and Bernard, Denis},
  journal = {Phys. Rev. X},
  volume = {13},
  issue = {1},
  pages = {011045},
  numpages = {28},
  year = {2023},
  month = {Mar},
  publisher = {American Physical Society},
  doi = {10.1103/PhysRevX.13.011045},
  url = {https://link.aps.org/doi/10.1103/PhysRevX.13.011045}
}

@article{Hruza_2023_entanglementlatetimelimitQSSEP,
	title = {Exact entanglement in the driven quantum symmetric simple exclusion process},
	pages = {175},
	author = {Bernard, Denis and Hruza, Ludwig},
	journal = {SciPost Phys.},
	volume = {15},
	year = {2023},
	publisher = {SciPost},
	doi = {10.21468/SciPostPhys.15.4.175},
	url = {https://scipost.org/10.21468/SciPostPhys.15.4.175}
}

@misc{barraquand_2025_introductionquantumexclusionprocesses,
      title={Introduction to quantum exclusion processes}, 
      author={Guillaume Barraquand and Denis Bernard},
      year={2025},
      eprint={2507.01570},
      archivePrefix={arXiv},
      primaryClass={math.PR},
      url={https://arxiv.org/abs/2507.01570}, 
}

@article{JoaoCosta_2026,
  title = {Emergence of Universality in Transport of Noisy Free Fermions},
  author = {Costa, Jo\~ao and Ribeiro, Pedro and De Luca, Andrea},
  journal = {Phys. Rev. Lett.},
  volume = {137},
  issue = {2},
  pages = {020403},
  numpages = {10},
  year = {2026},
  month = {Jul},
  publisher = {American Physical Society},
  doi = {10.1103/v8x8-ft81},
  url = {https://link.aps.org/doi/10.1103/v8x8-ft81}
}

@misc{StefanoScopa_2026_universalclassicalquantumfluctuations,
      title={Universal classical and quantum fluctuations in the large deviations of current of noisy quantum systems: The case of QSSEP and QSSIP}, 
      author={Mathias Albert and Denis Bernard and Tony Jin and Stefano Scopa and Shiyi Wei},
      year={2026},
      eprint={2601.16883},
      archivePrefix={arXiv},
      primaryClass={cond-mat.stat-mech},
      url={https://arxiv.org/abs/2601.16883}, 
}

@article{Alba_nuQSSEP_intro,
   title={Extension of quantum symmetric simple exclusion process model for entanglement spreading in stochastic diffusive quantum systems},
   volume={113},
   ISSN={2469-9969},
   url={http://dx.doi.org/10.1103/kbgw-lrfs},
   DOI={10.1103/kbgw-lrfs},
   number={7},
   journal={Physical Review B},
   publisher={American Physical Society (APS)},
   author={Alba, Vincenzo},
   year={2026},
   month=Feb }

@article{Alba_nuQSSEP_mpemba,
   title={Dynamics of entanglement fluctuations and quantum Mpemba effect in the $\nu = 1$ QSSEP model},
   volume={2026},
   ISSN={1742-5468},
   url={http://dx.doi.org/10.1088/1742-5468/ae4bb9},
   DOI={10.1088/1742-5468/ae4bb9},
   number={3},
   journal={Journal of Statistical Mechanics: Theory and Experiment},
   publisher={IOP Publishing},
   author={Russotto, Angelo and Ares, Filiberto and Calabrese, Pasquale and Alba, Vincenzo},
   year={2026},
   month=Mar, pages={033103} }

@misc{Alba_nuQSSEP_GHD,
      title={Inhomogeneous quenches and GHD in the $\nu = 1$ QSSEP model},
      author={Angelo Russotto and Filiberto Ares and Pasquale Calabrese and Vincenzo Alba},
      year={2026},
      eprint={2602.15122},
      archivePrefix={arXiv},
      primaryClass={cond-mat.stat-mech},
      url={https://arxiv.org/abs/2602.15122},
}

@article{Nahum_2025_random_majorana,
  title = {Spacetime picture for entanglement generation in noisy fermion chains},
  author = {Swann, Tobias and Bernard, Denis and Nahum, Adam},
  journal = {Phys. Rev. B},
  volume = {112},
  issue = {6},
  pages = {064301},
  numpages = {23},
  year = {2025},
  month = {Aug},
  publisher = {American Physical Society},
  doi = {10.1103/PhysRevB.112.064301},
  url = {https://link.aps.org/doi/10.1103/PhysRevB.112.064301}
}

@article{Nahum_2017_randomcircuitsentanglementgrowth,
  title = {Quantum Entanglement Growth under Random Unitary Dynamics},
  author = {Nahum, Adam and Ruhman, Jonathan and Vijay, Sagar and Haah, Jeongwan},
  journal = {Phys. Rev. X},
  volume = {7},
  issue = {3},
  pages = {031016},
  numpages = {30},
  year = {2017},
  month = {Jul},
  publisher = {American Physical Society},
  doi = {10.1103/PhysRevX.7.031016},
  url = {https://link.aps.org/doi/10.1103/PhysRevX.7.031016}
}

@article{Nahum_2018_randomcircuitsoperatorspreading,
  title = {Operator Spreading in Random Unitary Circuits},
  author = {Nahum, Adam and Vijay, Sagar and Haah, Jeongwan},
  journal = {Phys. Rev. X},
  volume = {8},
  issue = {2},
  pages = {021014},
  numpages = {30},
  year = {2018},
  month = {Apr},
  publisher = {American Physical Society},
  doi = {10.1103/PhysRevX.8.021014},
  url = {https://link.aps.org/doi/10.1103/PhysRevX.8.021014}
}

@article{Zhou_2020_randomcircuitsentanglementmembrane,
  title = {Entanglement Membrane in Chaotic Many-Body Systems},
  author = {Zhou, Tianci and Nahum, Adam},
  journal = {Phys. Rev. X},
  volume = {10},
  issue = {3},
  pages = {031066},
  numpages = {37},
  year = {2020},
  month = {Sep},
  publisher = {American Physical Society},
  doi = {10.1103/PhysRevX.10.031066},
  url = {https://link.aps.org/doi/10.1103/PhysRevX.10.031066}
}

@article{Pollmann_2018_randomcircuits,
  title = {Operator Hydrodynamics, OTOCs, and Entanglement Growth in Systems without Conservation Laws},
  author = {von Keyserlingk, C. W. and Rakovszky, Tibor and Pollmann, Frank and Sondhi, S. L.},
  journal = {Phys. Rev. X},
  volume = {8},
  issue = {2},
  pages = {021013},
  numpages = {19},
  year = {2018},
  month = {Apr},
  publisher = {American Physical Society},
  doi = {10.1103/PhysRevX.8.021013},
  url = {https://link.aps.org/doi/10.1103/PhysRevX.8.021013}
}

@article{Pollmann_2018_randomcircuitsconservationlaws,
  title = {Diffusive Hydrodynamics of Out-of-Time-Ordered Correlators with Charge Conservation},
  author = {Rakovszky, Tibor and Pollmann, Frank and von Keyserlingk, C. W.},
  journal = {Phys. Rev. X},
  volume = {8},
  issue = {3},
  pages = {031058},
  numpages = {28},
  year = {2018},
  month = {Sep},
  publisher = {American Physical Society},
  doi = {10.1103/PhysRevX.8.031058},
  url = {https://link.aps.org/doi/10.1103/PhysRevX.8.031058}
}

@article{Rakovszky_2019_randomcircuitsconservationlaws,
  title = {Sub-ballistic Growth of R\'enyi Entropies due to Diffusion},
  author = {Rakovszky, Tibor and Pollmann, Frank and von Keyserlingk, C. W.},
  journal = {Phys. Rev. Lett.},
  volume = {122},
  issue = {25},
  pages = {250602},
  numpages = {6},
  year = {2019},
  month = {Jun},
  publisher = {American Physical Society},
  doi = {10.1103/PhysRevLett.122.250602},
  url = {https://link.aps.org/doi/10.1103/PhysRevLett.122.250602}
}

@misc{Ribeirp2021,
      title={Diffusive Operator Spreading for Random Unitary Free Fermion Circuits}, 
      author={Beatriz C. Dias and Masudul Haque and Pedro Ribeiro and Paul McClarty},
      year={2021},
      eprint={2102.09846},
      archivePrefix={arXiv},
      primaryClass={cond-mat.str-el},
      url={https://arxiv.org/abs/2102.09846}, 
}

@misc{lastres2026geometryfreefermioncommutants,
      title={Geometry of Free Fermion Commutants}, 
      author={Marco Lastres and Sanjay Moudgalya},
      year={2026},
      eprint={2604.05031},
      archivePrefix={arXiv},
      primaryClass={quant-ph},
      url={https://arxiv.org/abs/2604.05031}, 
}

@article{Derrida_2009,
   title={Current Fluctuations of the One Dimensional Symmetric Simple Exclusion Process with Step Initial Condition},
   volume={136},
   ISSN={1572-9613},
   url={http://dx.doi.org/10.1007/s10955-009-9772-7},
   DOI={10.1007/s10955-009-9772-7},
   number={1},
   journal={Journal of Statistical Physics},
   publisher={Springer Science and Business Media LLC},
   author={Derrida, Bernard and Gerschenfeld, Antoine},
   year={2009},
   month=jun, pages={1--15} }

@article{Collins_2006,
   title={Integration with Respect to the Haar Measure on Unitary, Orthogonal and Symplectic Group},
   volume={264},
   ISSN={1432-0916},
   url={http://dx.doi.org/10.1007/s00220-006-1554-3},
   DOI={10.1007/s00220-006-1554-3},
   number={3},
   journal={Communications in Mathematical Physics},
   publisher={Springer Science and Business Media LLC},
   author={Collins, Benoit and Sniady, Piotr},
   year={2006},
   month=Mar, pages={773--795} }

@article{Mele_2024,
   title={Introduction to Haar Measure Tools in Quantum Information A Beginner's Tutorial},
   volume={8},
   ISSN={2521-327X},
   url={http://dx.doi.org/10.22331/q-2024-05-08-1340},
   DOI={10.22331/q-2024-05-08-1340},
   journal={Quantum},
   publisher={Verein zur Forderung des Open Access Publizierens in den Quantenwissenschaften},
   author={Mele, Antonio Anna},
   year={2024},
   month=May, pages={1340} }

@misc{mezzadri2007,
      title={How to generate random matrices from the classical compact groups}, 
      author={Francesco Mezzadri},
      year={2007},
      eprint={math-ph/0609050},
      archivePrefix={arXiv},
      primaryClass={math-ph},
      url={https://arxiv.org/abs/math-ph/0609050}, 
}

@article{Peschel_2003,
   title={Calculation of reduced density matrices from correlation functions},
   volume={36},
   DOI={10.1088/0305-4470/36/14/101},
   number={14},
   journal={Journal of Physics A: Mathematical and General},
   publisher={IOP Publishing},
   author={Peschel, Ingo},
   year={2003},
   month=Mar, pages={L205--L208} }

@article{Peschel_2009,
   title={Reduced density matrices and entanglement entropy in free lattice models},
   volume={42},
   DOI={10.1088/1751-8113/42/50/504003},
   number={50},
   journal={Journal of Physics A: Mathematical and Theoretical},
   publisher={IOP Publishing},
   author={Peschel, Ingo and Eisler, Viktor},
   year={2009},
   month=Dec, pages={504003} }

@ARTICLE{Levitov_1993,
       author = {{Levitov}, L.~S. and {Lesovik}, G.~B.},
        title = "{Charge distribution in quantum shot noise}",
      journal = {Soviet Journal of Experimental and Theoretical Physics Letters},
         year = 1993,
        month = aug,
       volume = {58},
        pages = {230},
       adsurl = {https://ui.adsabs.harvard.edu/abs/1993JETPL..58..230L}
}

@misc{klich_2002,
      title={Full Counting Statistics: An elementary derivation of Levitov's formula}, 
      author={I. Klich},
      year={2002},
      eprint={cond-mat/0209642},
      archivePrefix={arXiv},
      primaryClass={cond-mat.mes-hall},
      url={https://arxiv.org/abs/cond-mat/0209642}, 
}

@article{klich_levitov_2009,
  title = {Quantum Noise as an Entanglement Meter},
  author = {Klich, Israel and Levitov, Leonid},
  journal = {Phys. Rev. Lett.},
  volume = {102},
  issue = {10},
  pages = {100502},
  numpages = {4},
  year = {2009},
  month = {Mar},
  publisher = {American Physical Society},
  doi = {10.1103/PhysRevLett.102.100502},
  url = {https://link.aps.org/doi/10.1103/PhysRevLett.102.100502}
}

@article{Song_2012,
   title={Bipartite fluctuations as a probe of many-body entanglement},
   volume={85},
   DOI={10.1103/PhysRevB.85.035409},
   number={3},
   journal={Physical Review B},
   publisher={American Physical Society (APS)},
   author={Song, H. Francis and Rachel, Stephan and Flindt, Christian and Klich, Israel and Laflorencie, Nicolas and Le Hur, Karyn},
   year={2012},
   month=Jan, pages={035409} }

@article{Tagliacozzo_2022_review,
	title = {Fermionic Gaussian states: an introduction to numerical approaches},
	pages = {54},
	author = {Surace, Jacopo and Tagliacozzo, Luca},
	journal = {SciPost Phys. Lect. Notes},
	year = {2022},
	publisher = {SciPost},
	doi = {10.21468/SciPostPhysLectNotes.54},
	url = {https://scipost.org/10.21468/SciPostPhysLectNotes.54}
}

@article{Derrida_lectures_notes,
	title = {Lecture notes on large deviations in non-equilibrium diffusive systems},
	pages = {106},
	author = {Derrida, Bernard},
	journal = {SciPost Phys. Lect. Notes},
	year = {2025},
	publisher = {SciPost},
	doi = {10.21468/SciPostPhysLectNotes.106},
	url = {https://scipost.org/10.21468/SciPostPhysLectNotes.106}
}

@article{2026_Essler_vectorization,
	title = {Linear response and exact hydrodynamic projections in Lindblad equations with decoupled Bogoliubov hierarchies},
	pages = {058},
	author = {Penc, Patrik and Essler, Fabian H. L.},
	journal = {SciPost Phys.},
	volume = {20},
	year = {2026},
	publisher = {SciPost},
	doi = {10.21468/SciPostPhys.20.2.058},
	url = {https://scipost.org/10.21468/SciPostPhys.20.2.058}
}

@article{IgloiSzatmariLin_2012,
  title = {Entanglement entropy dynamics of disordered quantum spin chains},
  author = {Igl{\'o}i, Ferenc and Szatm{\'a}ri, Zsolt and Lin, Yu-Cheng},
  journal = {Phys. Rev. B},
  volume = {85},
  issue = {9},
  pages = {094417},
  numpages = {8},
  year = {2012},
  month = {Mar},
  publisher = {American Physical Society},
  doi = {10.1103/PhysRevB.85.094417},
  url = {https://link.aps.org/doi/10.1103/PhysRevB.85.094417}
}

@article{Fischer_TFIM,
  title = {Critical behavior of random transverse-field Ising spin chains},
  author = {Fisher, Daniel S.},
  journal = {Phys. Rev. B},
  volume = {51},
  issue = {10},
  pages = {6411--6461},
  numpages = {0},
  year = {1995},
  month = {Mar},
  publisher = {American Physical Society},
  doi = {10.1103/PhysRevB.51.6411},
  url = {https://link.aps.org/doi/10.1103/PhysRevB.51.6411}
}

@article{Fisher_1994,
  title = {Random antiferromagnetic quantum spin chains},
  author = {Fisher, Daniel S.},
  journal = {Phys. Rev. B},
  volume = {50},
  issue = {6},
  pages = {3799--3821},
  numpages = {0},
  year = {1994},
  month = {Aug},
  publisher = {American Physical Society},
  doi = {10.1103/PhysRevB.50.3799},
  url = {https://link.aps.org/doi/10.1103/PhysRevB.50.3799}
}

@article{Evers_2008,
   title={Anderson transitions},
   volume={80},
   ISSN={1539-0756},
   url={http://dx.doi.org/10.1103/RevModPhys.80.1355},
   DOI={10.1103/revmodphys.80.1355},
   number={4},
   journal={Reviews of Modern Physics},
   publisher={American Physical Society (APS)},
   author={Evers, Ferdinand and Mirlin, Alexander D.},
   year={2008},
   month=Oct, pages={1355–1417} }

@article{Atland_Zirnbaurer,
  title = {Nonstandard symmetry classes in mesoscopic normal-superconducting hybrid structures},
  author = {Altland, Alexander and Zirnbauer, Martin R.},
  journal = {Phys. Rev. B},
  volume = {55},
  issue = {2},
  pages = {1142--1161},
  numpages = {0},
  year = {1997},
  month = {Jan},
  publisher = {American Physical Society},
  doi = {10.1103/PhysRevB.55.1142},
  url = {https://link.aps.org/doi/10.1103/PhysRevB.55.1142}
}

@article{Dyson_1953,
  title = {The Dynamics of a Disordered Linear Chain},
  author = {Dyson, Freeman J.},
  journal = {Phys. Rev.},
  volume = {92},
  issue = {6},
  pages = {1331--1338},
  numpages = {0},
  year = {1953},
  month = {Dec},
  publisher = {American Physical Society},
  doi = {10.1103/PhysRev.92.1331},
  url = {https://link.aps.org/doi/10.1103/PhysRev.92.1331}
}

@article{IgloiJuhasz_2008,
doi = {10.1209/0295-5075/81/57003},
url = {https://doi.org/10.1209/0295-5075/81/57003},
year = {2008},
month = {feb},
publisher = {},
volume = {81},
number = {5},
pages = {57003},
author = {Iglói, F. and Juhász, R.},
title = {Exact relationship between the entanglement entropies of XY and quantum Ising chains},
journal = {Europhysics Letters},
}

\end{document}